%% file: mobihoc_main.tex
\documentclass[sigconf, anonymous=false, review=false, screen, 10pt]{acmart} 

\usepackage{xcolor}
\usepackage{xurl}
\usepackage{booktabs}
\usepackage[ruled]{algorithm2e}
\usepackage{algpseudocode}
\usepackage{enumerate}
\usepackage[english]{babel}
\usepackage{blindtext}
\usepackage[caption=false,font=footnotesize,subrefformat=parens]{subfig}
\usepackage{amsmath}

\usepackage{amssymb}
\usepackage{xspace}
\usepackage{multirow}
\usepackage{threeparttable}
\usepackage{float}
\usepackage{flafter}
\usepackage{comment}
\usepackage[skip=12pt]{caption}
\usepackage{graphicx}
\usepackage{gensymb}

\def\fig{Fig.\xspace}
\def\eqn{Eq.\xspace}

\def\tab{Tab.\xspace}

\def\ie{{\textit{i.e.}\xspace}} 
\def\eg{{\textit{e.g.}\xspace}}
\def\aka{{\textit{a.k.a.}\xspace}} 

\def\etc{{\textit{etc}\xspace}}

\newcommand{\head}[1]{{\noindent \textbf{#1:}}}

\graphicspath{{./figures/}}
\graphicspath{{./pdf/}}
\DeclareGraphicsExtensions{.pdf,.jpeg,.png}

\ifodd 0
\newcommand{\rev}[1]{{\color{blue}#1}} 
\newcommand{\com}[1]{\textbf{\color{red}(COMMENT: #1)}} 
\newcommand{\todo}[1]{\textbf{{\color{orange}(TODO: #1)}}}
\newcommand{\unused}[1]{}
\else
\newcommand{\rev}[1]{#1}
\newcommand{\com}[1]{}
\newcommand{\todo}[1]{}
\newcommand{\unused}[1]{}
\fi

\setcopyright{acmcopyright}
\setcopyright{acmlicensed}

\copyrightyear{2024} 
\acmYear{2024} 
\setcopyright{acmlicensed}\acmConference[MOBIHOC '24]{The Twenty-fifth International Symposium on Theory, Algorithmic Foundations, and Protocol Design for Mobile Networks and Mobile Computing}{October 14--17, 2024}{Athens, Greece}
\acmBooktitle{The Twenty-fifth International Symposium on Theory, Algorithmic Foundations, and Protocol Design for Mobile Networks and Mobile Computing (MOBIHOC '24), October 14--17, 2024, Athens, Greece}
\acmDOI{10.1145/3641512.3686357}
\acmISBN{979-8-4007-0521-2/24/10}

\acmPrice{}

\def\sysname{\textsc{TADAR}\xspace}
\begin{document}

\title{\sysname: Thermal Array-based Detection and Ranging for Privacy-Preserving Human Sensing
}

\begin{anonsuppress}
	\author{Xie Zhang}
	\email{zhangxie@connect.hku.hk}
	\affiliation{%
		\institution{Department of Computer Science\\The University of Hong Kong}
		\country{Hong Kong}
	}
	
	\author{Chenshu Wu}
	\email{chenshu@cs.hku.hk}
	\affiliation{%
		\institution{Department of Computer Science\\The University of Hong Kong}
		\country{Hong Kong}
	}
\end{anonsuppress}

\renewcommand{\shortauthors}{Xie Zhang, Chenshu Wu}

\begin{abstract}
Human sensing has gained increasing attention in various applications. 
Among the available technologies, visual images offer high accuracy, while sensing on the RF spectrum preserves privacy, creating a conflict between imaging resolution and privacy preservation. 
In this paper, we explore thermal array sensors as an emerging modality that strikes an excellent resolution-privacy balance for ubiquitous sensing. 
To this end, we present \sysname, the first multi-user Thermal Array-based Detection and Ranging system that estimates the inherently missing range information, extending thermal array outputs from 2D thermal pixels to 3D depths and empowering them as a promising modality for ubiquitous privacy-preserving human sensing. 
We prototype \sysname using a single commodity thermal array sensor and conduct extensive experiments in different indoor environments. 
\rev{Our results show that \sysname achieves a mean F1 score of 88.8\% for multi-user detection and a mean accuracy of 32.0 cm for multi-user ranging, which further improves to 20.1 cm for targets located within 3 m.} 
We conduct two case studies on fall detection and occupancy estimation to showcase the potential applications of \sysname. 
We hope \sysname will inspire the vast community to explore new directions of thermal array sensing, beyond wireless and acoustic sensing. 
\textit{TADAR is open-sourced on GitHub:} \url{https://github.com/aiot-lab/TADAR}.

\end{abstract}



\maketitle 

\input{body_mobihoc}

\bibliographystyle{ACM-Reference-Format}
\bibliography{refs_mobihoc}

\end{document}

%% file: body_mobihoc.tex

\vspace{-0.1in}
\section{Introduction}
\label{sec:intro}

\begin{figure}[t]
    \centering
    \includegraphics[width=0.47\textwidth]{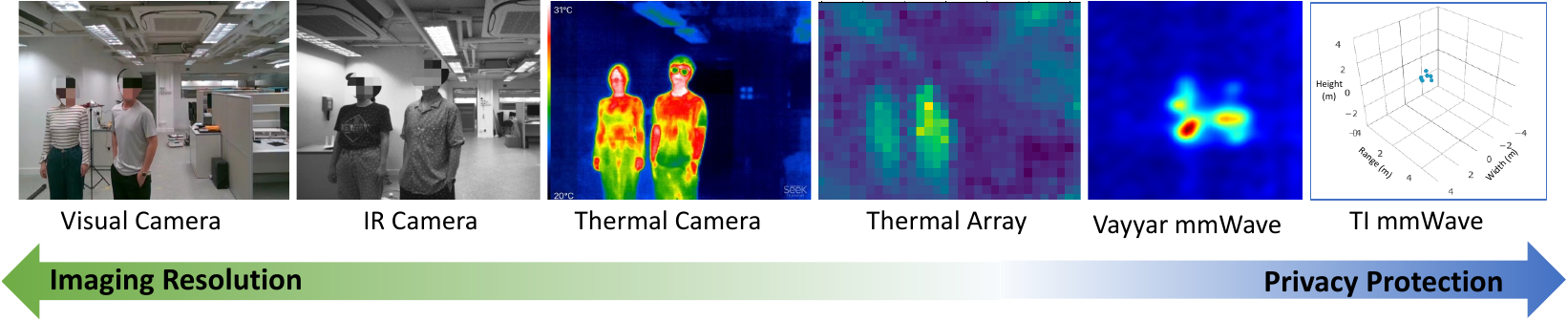}
    \vspace{-0.1in}
    \caption{Modality Comparison.}
    \label{fig:modality-images}
\end{figure}

Numerous applications are enabled by human-centric sensing, from people-aware environments to precision behavioral health, daily living monitoring, digital eldercare, and human-robot interactions, just to name a few. 
At the core of human sensing is the detection and ranging of human targets, which has been explored through various modalities like cameras \cite{xie2021ultradepth}, Sonar \cite{wang2016device}, Lidar \cite{taipalus2011human}, Radar \cite{zhang2020mmeye}, and Widar \cite{qianWidar2PassiveHuman2018}.
From a holistic view, however, there exists an inherent trade-off among various sensing modalities. 
In general, as portrayed in \fig\ref{fig:modality-images}, visible-spectrum sensing offers the best imaging resolution but raises serious privacy concerns, rendering visual cameras unfavorable in private spaces such as homes. 
In contrast, wireless sensing promises privacy preservation, yet at the cost of losing high resolution, making multi-user sensing a longstanding challenge \cite{xie2019md}. 
The future of non-contact human sensing calls for a new sensing modality compromising \textit{resolution} and \textit{privacy}. 

\begin{table}[tp]
\caption{Contact-less human sensing technologies}
\centering
\label{tab:modality_comparison}
\scriptsize
\vspace{-0.1in}
\begin{tabular}{ccccccl}
\cline{1-6}
\textbf{\begin{tabular}[c]{@{}c@{}}Sensing \\ Modality\end{tabular}} 
& \textbf{\begin{tabular}[c]{@{}c@{}}Imaging \\ Resolution\end{tabular}} 
& \textbf{\begin{tabular}[c]{@{}c@{}}Privacy \\ Protection\end{tabular}} 
& \textbf{\begin{tabular}[c]{@{}c@{}}Sensing \\ Mode\end{tabular}} 
& \textbf{Cost}   
&  \\ \cline{1-6}

RGB/IR Camera                   & Very High       & Very Weak                   & Quasi-passive & Low &  \\ 
Lidar               & Very High            & Weak                     & Active        & High & \\ 
Thermal Camera               & High            & Weak                       & Fully-passive & High &  \\ 
ToF Cameras               & High            & Weak                     & Active        & Medium & \\ 
\textbf{Thermal Array}       & \textbf{Medium}  & \textbf{Medium}   & \textbf{Fully-passive}  & \textbf{Low} &  \\ 
mmWave Radar                 &Low              & Medium                    & Active       & High  &  \\ 
Smart Speakers               & Low             & Weak                      & Active       & Low   &  \\ 
WiFi Devices                 & Low             & Strong                     & Active       & Low  &  \\ 
PIR Sensor                   & Very Low        & Strong                   &  Fully-passive & Low    & \\ \cline{1-6}
\end{tabular}
\end{table}

Recently, HADAR \cite{baoHeatassistedDetectionRanging2023} has emerged as a fully passive paradigm for Heat-Assisted Detection and Ranging, as opposed to active Sonar, Lidar, Radar, and quasi-passive RGB and Infrared (IR) cameras\footnote{While all being device-free, they are active sensing in the sense of actively transmitting certain signals or relying on external lights for sensing.}. 
However, HADAR, based on thermal cameras, faces privacy concerns and is prohibitively expensive for ubiquitous usage. 
To this end, we investigate \textit{Thermal Array-based Detection And Ranging} (TADAR) for fully passive, low-cost, and privacy-preserving human sensing. 
A thermal array sensor is an integrated array of thermopile detectors that operate within the Longwave Infrared (LWIR) spectrum. 
It features magnitude lower resolution and cost than thermal cameras\footnote{Thermal cameras often deliver hundreds of thousands of pixels (\eg, 640$\times$480, 1024$\times$768), offering a high enough clarity to infer privacy-concerned details and having a significantly higher price than thermal array sensors. 
For example, FLIR ONE Edge Pro thermal camera features 160$\times$120 pixels and costs about USD 600. 
In comparison, the Melexis MLX90641 thermal array sensor of 16$\times$12 pixels has a $\sim$15$\times$ lower bulk price.}. 
The low resolution significantly eliminate privacy concerns compared to common thermal cameras and RGB/IR cameras. 
On the other hand, the spatial resolution is still considerably high compared to wireless sensing using commercial WiFi \cite{pallaprolu2022wiffract} or compact millimeter-wave radars \cite{zhang2020mmeye}. 
As further detailed in \tab\ref{tab:modality_comparison}, thermal arrays offer a compelling balance between imaging resolution and privacy, rendering them an attractive solution to ubiquitous human sensing. 
In comparison, RGB cameras present substantial privacy risks that persist even with blurring and pixelation measures \cite{raviReviewVisualPrivacy2023}.
IR and ToF cameras operate within the near-infrared spectrum, which significantly differs from thermal array sensors, and have recently revealed concerns on user privacy \cite{xieMozartMobileToF2023}. 
Albeit thermal cameras also work on the LWIR spectrum, thermal array sensors emerge as a much lower-resolution, lower-cost, and more compact modality using distinct physical principles.

\begin{figure}[t]
    \centering
    \includegraphics[width=0.47\textwidth]{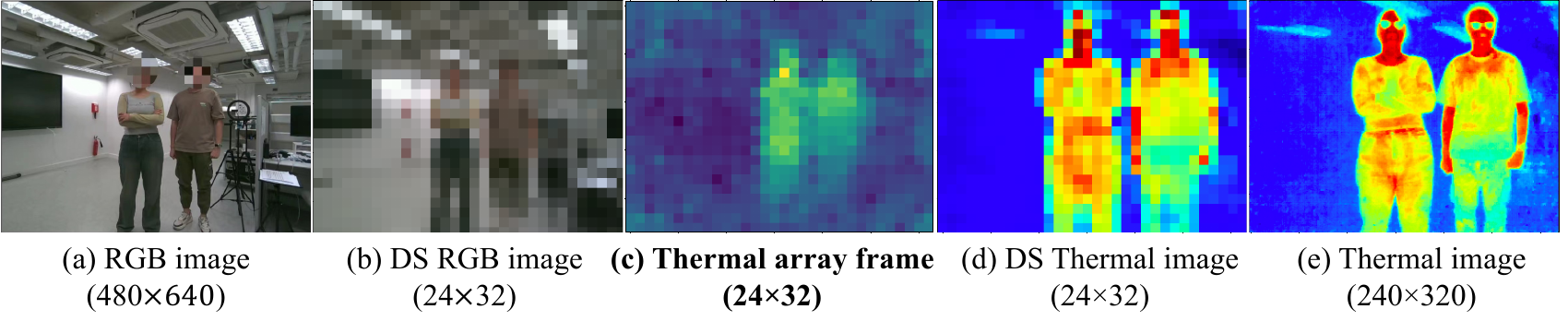}
    \vspace{-0.1in}
     \caption{Comparison of (down-sampled, DS) RGB/thermal images and thermal array frames. }
    \label{fig:rgb_thermal_thermal array}
\end{figure}

Like cameras, unfortunately, thermal array sensors are inherently missing the range information and only generate a 2D temperature map of objects inside its field of view (FOV). 
Range, \ie, the distance from a sensor to a target, is a key enabler for many sensing applications and also a major advantage of wireless sensing over visual perception. 
Without range, the scant literature on thermal array-based sensing is limited to using 2D thermal pixels as a regular image
\cite{naserMultipleThermalSensor2022, chiduralaDetectionMovingObjects2022,tatenoPrivacyPreservedFallDetection2020}, forcing sensors to be installed on ceilings\cite{chiduralaDetectionMovingObjects2022, chiduralaOccupancyEstimationUsing2021}. 
Therefore, we ask the question: \textit{Can we enable accurate ranging for thermal array-based sensing?} 

We present \sysname, a thermal array-based multi-user detection and ranging system that affirmatively answers the above question. 
By converting 2D thermal pixels of human targets into 3D coordinates with range, \sysname extends thermal array sensors for 3D spatial sensing just like mmWave radars and depth cameras while enjoying a good balance of utility and privacy. 
Our key insight is that, as the air attenuates the thermal radiation, human body temperatures detected at the thermal array sensor vary across different sensor-human distances (\ie, ranges). 
Notably, the ranging principle differs from that of the HADAR \cite{baoHeatassistedDetectionRanging2023}, which resorts to stereovision.
Intuitively, a thermal array sensor will capture a higher temperature for a closer human subject while a lower reading for a more distant one. 
Transforming such an intuition into a functional system poses significant challenges. 
First, the temperature readings of the human body depend on many factors besides distance, such as clothing, orientation, pose, sensor noise, \etc.
Additionally, for the same subject and thus the same range, the detected temperatures also differ across body parts. 
Second, the extremely low resolution of thermal array sensors prevents reliable temperature measurements of distant subjects, which can be confused with the background temperature or dominated by noise. 
Third, the combination of low resolution, textureless reading, and noise also makes it difficult to detect and isolate the thermal pixels from multiple targets.
\fig\ref{fig:rgb_thermal_thermal array} partly illustrates these issues with real measurements using MLX90640 thermal array sensor, rendering \sysname a unique problem different from computer vision based on RGB/thermal images. 

\sysname overcomes these challenges and delivers a practical multi-user detection and ranging system through two distinct designs. 
First, based on an in-depth understanding of thermal array sensor principles, we propose a novel mechanism that converts body temperatures into ranges. 
\rev{Our approach fully incorporates temperature readings from all body parts while prioritizing the more reliable and identifiable facial and neck temperatures as the primary references to predict the range, which allows it to maintain robustness across different body sizes and partial blockages}. 
Second, to support multi-user ranging, we present an efficient approach that detects and segments multiple users from textureless, low-resolution, noisy thermal array readings without training. 
The problem is very different from learning-based detection and segmentation on RGB images as evidenced by our experiments in \S\ref{sec:exp}. 
\sysname distinctively exploits the spatial distribution and temporal continuity of thermal array readings to determine the number of users and locate their regions accurately and robustly under various settings.   

We implement a real-time prototype of \sysname using a single commodity sensor and extensively evaluate \sysname in various environments. 
\rev{Our results show that \sysname achieves a mean F1 score of 88.8\% for multi-user detection and a mean accuracy of 0.32 m for multi-user ranging, improving to 0.20 m for targets located within 3 m.}
By enabling accurate ranging, many sensing applications can be explored in realistic scenarios with thermal array sensors easily mounted to the wall or placed on the table, such as location tracking and daily activity monitoring. 
To show this, we conduct two case studies for fall detection and occupancy monitoring. 
We believe that \sysname paves the way for ubiquitous thermal array sensing and will inspire new directions for research in human sensing. 

\head{Contributions}
Our contributions are summarized below. 
We explore and exploit thermal array sensors as a new and attractive modality for ubiquitous human sensing, which offers an excellent resolution-privacy balance. 
\rev{We present \sysname that enables the inherently unavailable range information for thermal array sensing, even with \textit{multiple users}.}
We build a real-time prototype system using commodity thermal array sensors and conduct extensive experiments with two case studies to validate the performance. 
To the best of our knowledge, \sysname is the first thermal array-based multi-user detection and ranging system, which can open up new directions for ubiquitous human sensing.

\vspace{-0.1in}
\section{thermal array Ranging}
\label{sec:ranging}

\subsection{Principle and Measurements}
A thermal array sensor, which is composed of thermopile detectors arranged in a grid, generates a 2D temperature map of objects within its FOV, as depicted in \fig\ref{fig:rgb_thermal_thermal array}.
\rev{The MLX90640BAA sensor used in \sysname has 32$\times$24 pixels, a $75^{\circ} \times 110^{\circ}$ FOV, and a reported temperature precision of $\pm 1.5^{\circ}C$, which is outside the clinically relevant limit of $\pm 0.5^{\circ}C$ \cite{simpson2019prospective}. Therefore, it is unlikely to compromise user privacy regarding temperature information}.
We further provide a brief overview of the thermography theory that underlies the thermal array sensors and conduct real-world measurements to demonstrate the feasibility and difficulties of thermal array ranging. 

\head{Thermography}
Infrared thermal imaging, \aka~thermography \cite{mollmannInfraredThermalImaging2018}, generates thermal images of objects by detecting their thermal radiation. 
Assuming lossless radiation, the object temperature is estimated by solving  \cite{mollmannInfraredThermalImaging2018}: 
\begin{equation}
    \epsilon_{obj} \cdot \int_{\lambda_1}^{\lambda_2} M(T_{obj}) d \lambda =  \Phi_{obj},
    \label{eq:temperature-measure}
\end{equation}
where ${\Phi}_{obj}$ is the total radiant power of the object in the spectral range $[\lambda_1, \lambda_2]$. $T_{obj}$ denotes the surface temperature of the object. $M(T)$ represents the unit radiant power of a blackbody at a temperature $T$, which can be calculated based on Planck's law.
$\epsilon_{obj}$ denotes the emissivity, the ratio of the amount of radiant power emitted from the object's surface to that emitted by the blackbody at the same temperature, which is assumed to be independent of wavelength and temperature \cite{diakides2012medical}.

In practice, the thermal array sensors detect the radiant power at the destination, \ie, the sensor itself, where the radiant power attenuates through the propagation medium, following the Bouguer–Lambert-Beer law \cite{mayerhoferBouguerBeerLambert2020}.
The actual temperature estimation, therefore, is described as
 \begin{equation}
    \int_{\lambda_1}^{\lambda_2} M(\hat{T}_{obj}) d \lambda = \exp{(-\gamma (\lambda) \cdot r)} \cdot \Phi_{obj},
    \label{eq:real-temperature-measure}
\end{equation}
where $\hat{T}_{obj}$ is the estimated temperature of the object at the sensor and $\gamma (\lambda)$ is the attenuation coefficient. $r$ is the traveled distance, \ie, the range between the object and the sensor.

\begin{figure*}[t]
  \begin{minipage}{0.24\textwidth}
    \centering
    \includegraphics[width=1.0\textwidth]{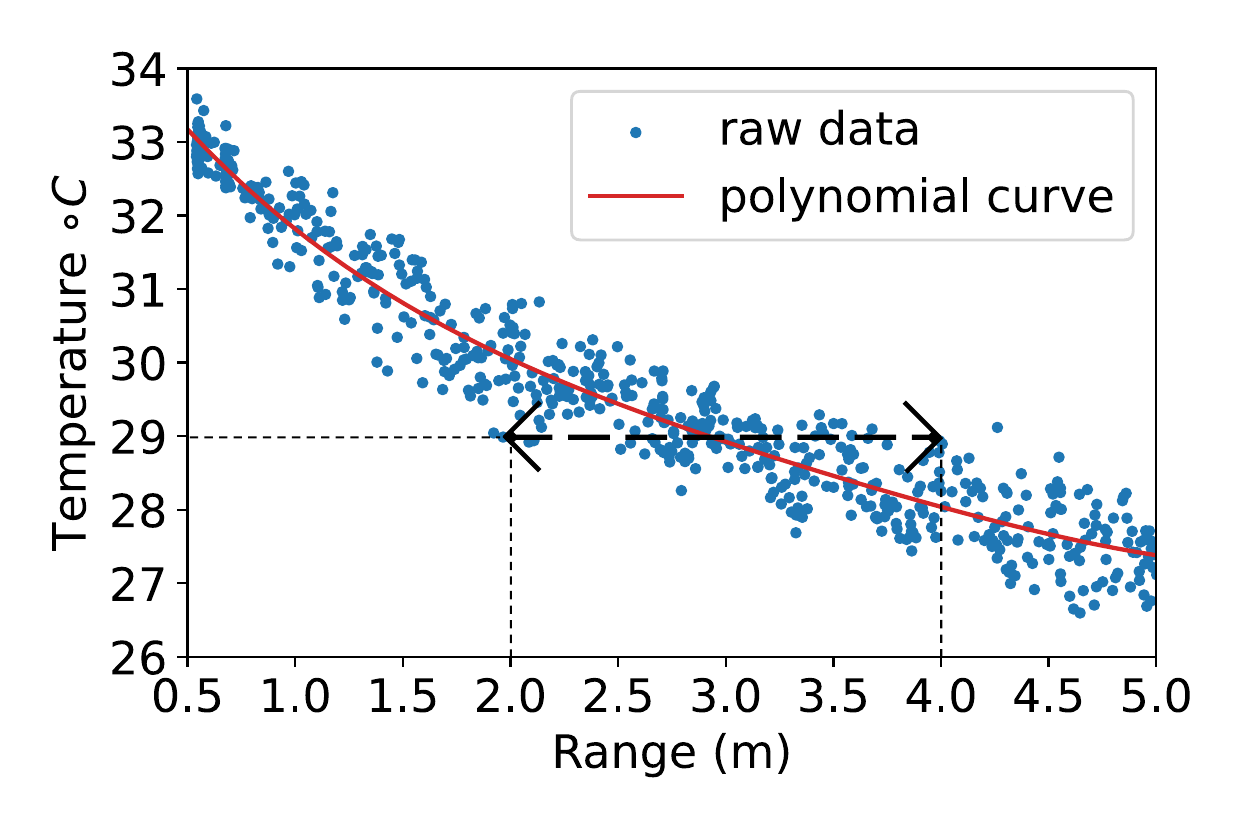}
    \vspace{-0.3in}
    \caption{Range and temperature relation. 
    }
    \label{fig:feasibility-human}
  \end{minipage}
  \hfill
  \begin{minipage}{0.43\textwidth}
    \centering
    \includegraphics[width=1.0\textwidth]{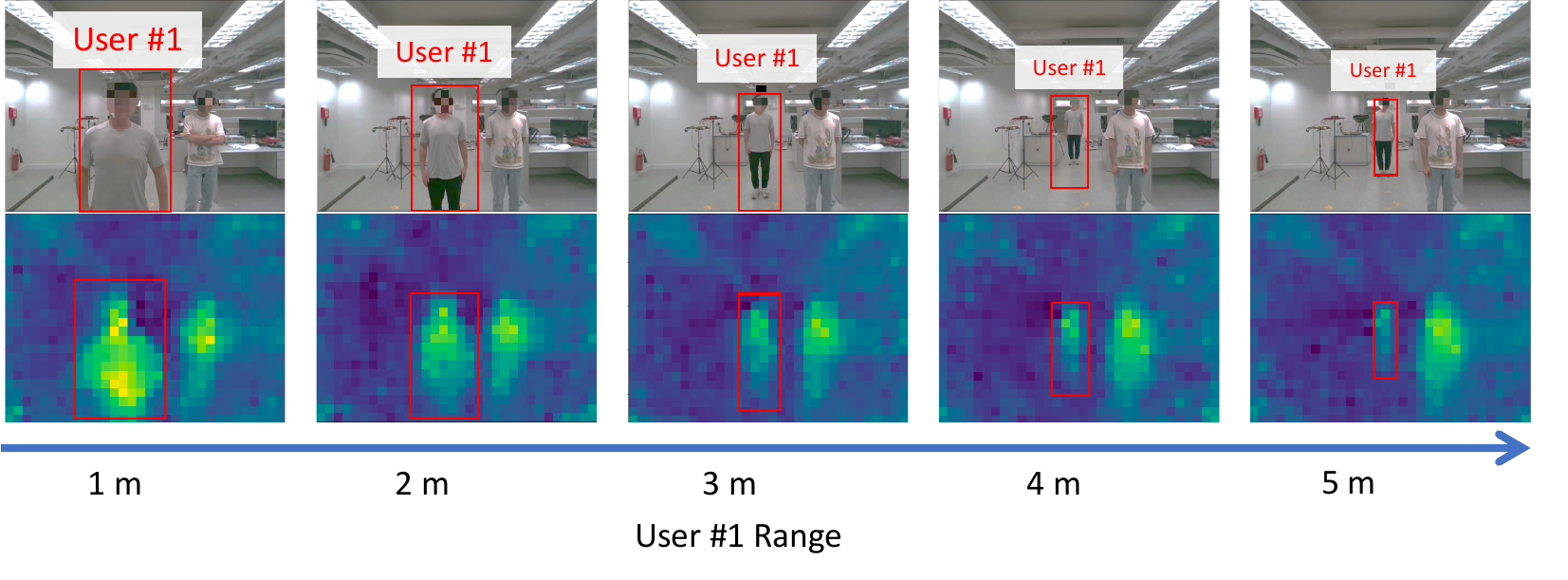}
    \vspace{-0.3in}
    \caption{Range and temperature map relation. 
    }
    \label{fig:range-temperature}
  \end{minipage}
  \hfill
  \begin{minipage}{0.29\textwidth}
    \centering
    \includegraphics[width=1.0\textwidth]{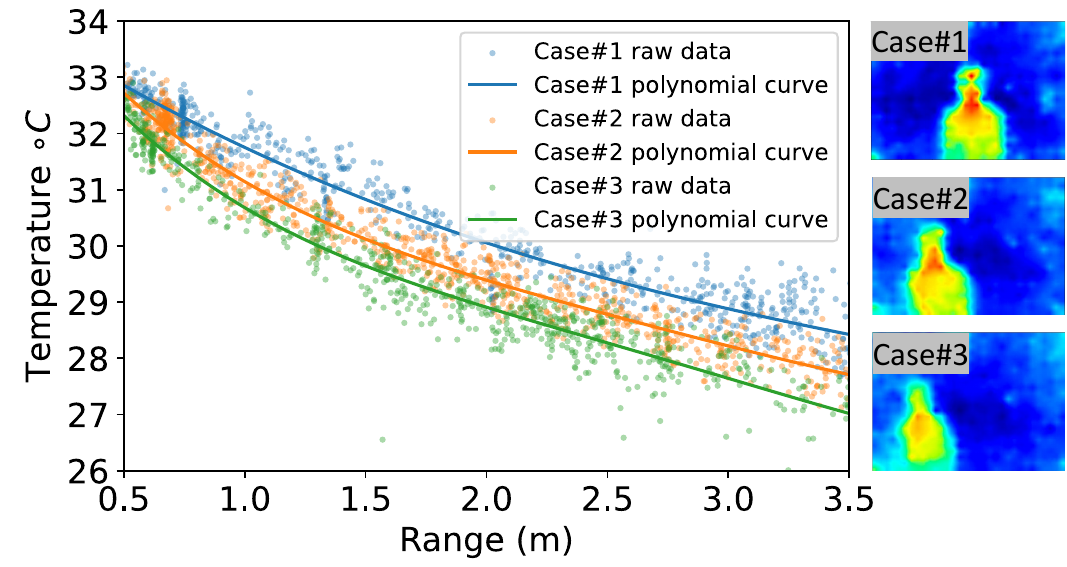}
    \vspace{-0.3in}
    \caption{The sensor's horizontal bias effect.}
    \label{fig:postprocess-range}
  \end{minipage}
\end{figure*}

\head{From Temperature to Range}
\eqn\eqref{eq:real-temperature-measure} sheds light on a principled approach to range estimation using the measured temperature $\hat{T}_{obj}$. 
Specifically, the relation between the detected temperature $\hat{T}_{obj}$ and the sensor-object distance $r$ can be established as:
 \begin{equation}
     r = \frac{-1}{\gamma (\lambda)} \ln{\frac{\int_{\lambda_1}^{\lambda_2} M(\hat{T}_{obj}) d \lambda}{\epsilon_{obj} \int_{\lambda_1}^{\lambda_2} M(T_{obj}) d \lambda}}.
    \label{eq:range-temperature}
\end{equation}
In theory, given that $\hat{T}_{obj}$ is reported by the sensor and assuming the actual temperature $T_{obj}$ of daily objects (\eg, the human body) is known, we can directly calculate the range $r$ based on the above equation, if $\gamma (\lambda)$ is also available.

This ideal model, however, is infeasible to apply to realistic thermal array measurements in practice. 
First, for human targets, the detected temperature $\hat{T}_{user}$ depends on various factors besides the actual body temperature $T_{user}$ and the range $r$. 
Second, the spectral range $[\lambda_1, \lambda_2]$ is determined by hardware properties of the thermal array and is not publicly available. 
Third, the temperature readings by the thermal array are also affected by hardware noise and the background environments. 
Consequently, we cannot analytically solve the above theoretical model to derive the range $r$.

\vspace{-0.1in}
\subsection{Thermal Array-based Human Ranging}

Targeting human sensing applications, we focus on human subjects and build a practical approach for human ranging. 
Our design is motivated by three observations: 
First, the above analysis inspires us that there does exist a certain mapping relationship between the sensor-reported temperatures and the range. As shown in \fig\ref{fig:feasibility-human}, the detected highest temperatures of the user's body overall decrease monotonically and non-linearly, with respect to the range.
Second, people always expose certain areas of bare skin, typically the face and neck areas, to the air in indoor spaces. 
It promises an opportunity to employ facial temperatures for robust ranging, since the skin emissivity is almost stable at 0.98, and the difference in the facial temperature among different persons is less than $0.7^\circ C$ \cite{christensenThermographicImagingFacial2012}, which is negligible compared to the impact of other factors. 
Third,  as shown in \fig\ref{fig:range-temperature}, the on-body pixels except for the user's face and neck areas also provide valuable range information that may boost the ranging performance.
These insights inspire us to build a ranging model to predict the distance using the user's temperature readings. 

We first assume that the region of interest (ROI) in the thermal array map has been detected for a target user, and leave thermal target detection to the next section \S\ref{sec:design}. 
A naive method for range estimation is to construct a regression model $R(\cdot)$ that approximates the relation in \eqn\ref{eq:range-temperature} and produces an estimated range $\hat{r} = R(T_{(x^{\prime},y^{\prime})})$, where $T_{(x^{\prime},y^{\prime})} \in \mathcal{ROI}$ is the selected representative temperature in the target's ROI. However, relying on a single pixel value due to sensor noise can lead to large errors, as shown in \fig\ref{fig:feasibility-human}.

\head{Thermal ranging feature extraction} 
To construct a more reliable and accurate range estimation model, we consider all temperature pixels inside the target's region to extract the ranging feature. 
However, effectively integrating all information in the region requires addressing three challenges: First, the region size is related to the user's body size, which may cause bias in the range estimation model. 
Second, different postures and body orientations can cause varying temperature distributions inside the 2D region, leading to unstable model performance. Third, the user's clothing can alter the detected temperature inside the ROI, posing a challenge to the robustness of range estimation.

To mitigate the bias caused by body size and maximize information utilization, we employ the ROI pooling technique \cite{renFasterRCNNRealTime2015a}, denoted as $Pool(\cdot)$, to extract features from all on-body pixels. Specifically, the ROI is divided into fixed rectangular cells, with each cell's maximum temperature serving as a feature element. This approach guarantees that the representations of the ROIs have uniform shapes, thereby eliminating any dependence on body size.
Furthermore, to reduce the impact of user postures and orientations, the pooled ROI is flattened $F(\cdot)$ and sorted $Sort(\cdot)$, forming a temperature feature vector $\mathbf{\tau} \in \mathcal{R}^{n}$. This process eliminates the 2D temperature distribution, resulting in an unbiased feature representation. Lastly, to avoid the impact of the user's clothing, we rely on pixels of facial and neck temperatures in the ROI. \rev{{Luckily, we notice that the naked facial and neck areas always touch the highest readings in the ROI because any other covered body parts will suffer from larger attenuation or be completely blocked, resulting in lower temperatures observed at the sensor}}. Therefore, we primarily rely on pixels with higher temperatures for range estimation, which circumvents the complex impact of clothing. Notably, the ROI pooling and flattening operations preserve the temperatures of the facial and neck areas. Additionally, the sorting operation arranges the higher temperature values at the beginning of the feature vector, hereby prioritizing these areas for the ranging model.

\begin{figure}[t]
    \centering
    \includegraphics[width=0.45\textwidth]{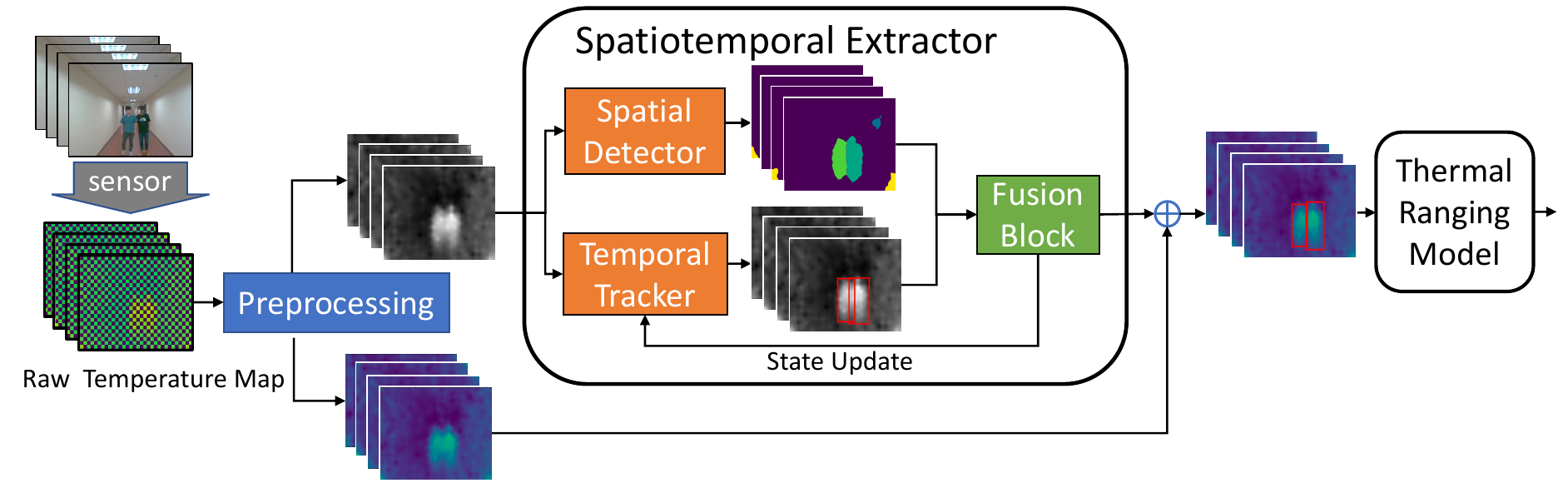}
    \caption{\sysname structure. 
    }
    \vspace{-0.1in}
    \label{fig:infrage-overview}
\end{figure}

\head{Thermal ranging compensation} 
\rev{As depicted in \fig\ref{fig:postprocess-range}, the detected body temperatures can be unstable, depending on the target's relative position within the frame.}
Specifically, the temperature readings are generally lower when a target is closer to the boundary of the thermal array frame, \ie, horizontal bias.
To combat horizontal bias, we introduce a simple yet effective approach that incorporates the center point location of the ROI into the temperature feature vector $\tau$, serving as a compensation feature to construct the final thermal ranging feature $\mathbf{\tau}'$, \ie, $\mathbf{\tau}' = \mathbf{\tau} \oplus (x_{center}, y_{center})$. 
The efficacy of this technique is substantiated through experiments (\S\ref{ssec:exp_be}).

\head{Thermal ranging model}
We then investigate a regression model, $R': \mathbf{\tau}' \to \hat{r}$, to estimate the range $\hat{r}$ from the thermal ranging feature $\mathbf{\tau}'$. 
To keep the model lightweight and readily deployable on embedded devices, we mainly explore conventional machine learning models and leave deep neural networks for the future. 
We have considered different models \cite{murphy2012machine}, such as Hist Gradient Boosting (HGB), Kernel Ridge Regression (KRR), and Support Vector Regression (SVR). 
In \sysname, we adopt HGB as the backbone in our thermal ranging model, which demonstrates excellent performance among others. Additionally, we consider the temporal continuity of the target's range and utilize the Kalman filter over the range estimations from consecutive thermal array maps to enhance the ranging accuracy as demonstrated in \S\ref{ssec:exp_be}. 
\rev{The final form of our thermal ranging model is:
\begin{equation}
    \hat{r} = R'( Sort(F(Pool(T_{(x^{\prime}, y^{\prime})}))) \oplus (x_{center}, y_{center}))
    \label{eq:range_model_eq}
\end{equation}}
where $T_{(x^{\prime}, y^{\prime})} \in \mathcal{ROI}$ and $\hat{r}$ is the estimated range.

\section{\sysname Design}
\label{sec:design}
This section presents the system design of \sysname, as illustrated in \fig\ref{fig:infrage-overview}. 
Importantly, we propose an effective and efficient approach to thermal array-based multi-user detection, which is assumed to be available in the previous section. 

\begin{figure*}[t]
  \begin{minipage}{0.32\textwidth}
    \centering
    \includegraphics[width=1.0\textwidth]{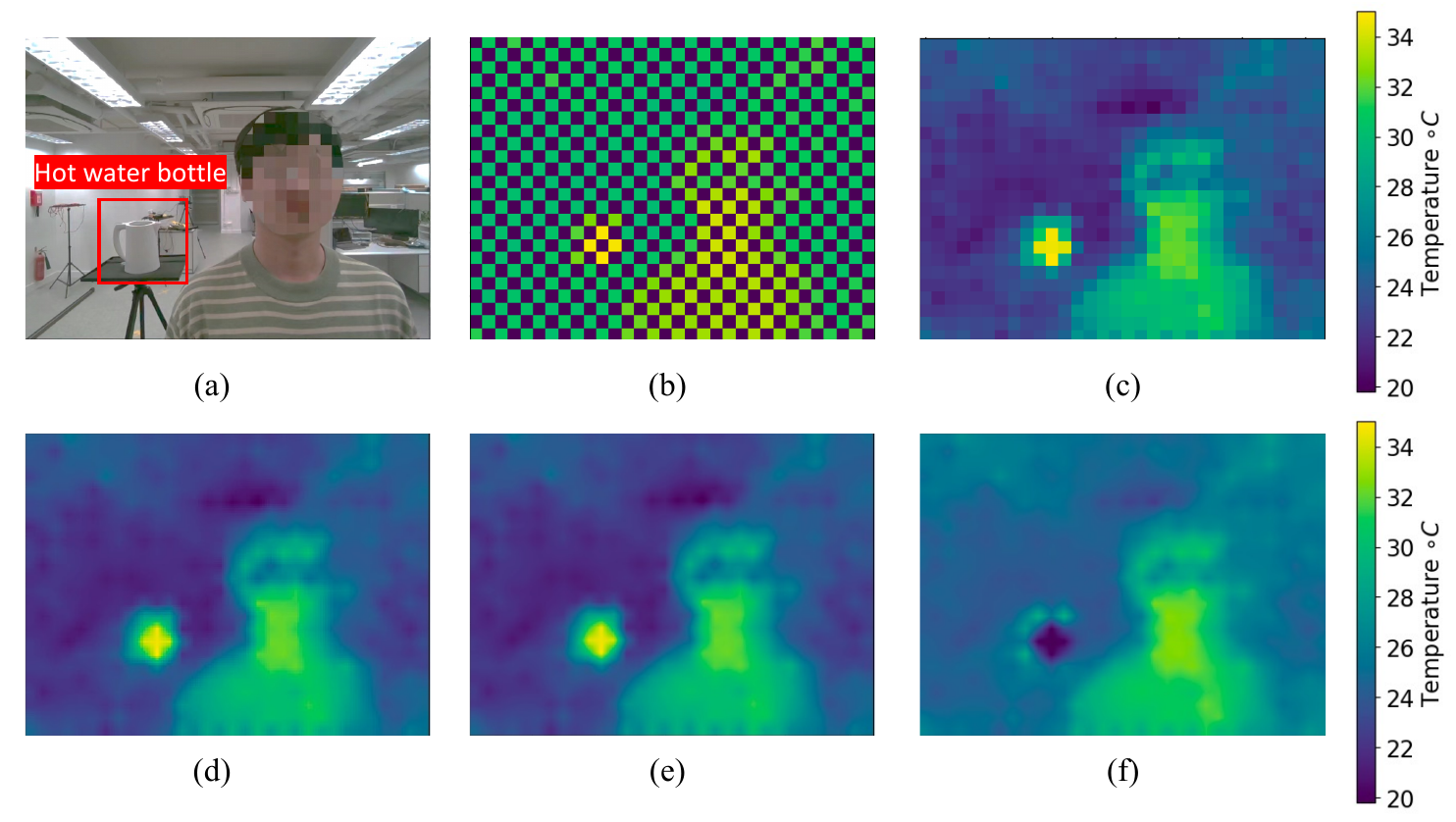}
    \vspace{-0.3in}
    \caption{Preprocssing. 
    }
    \label{fig:preprocessing}
  \end{minipage}
  \hfill
  \begin{minipage}{0.3\textwidth}
    \centering
    \includegraphics[width=1.0\textwidth]{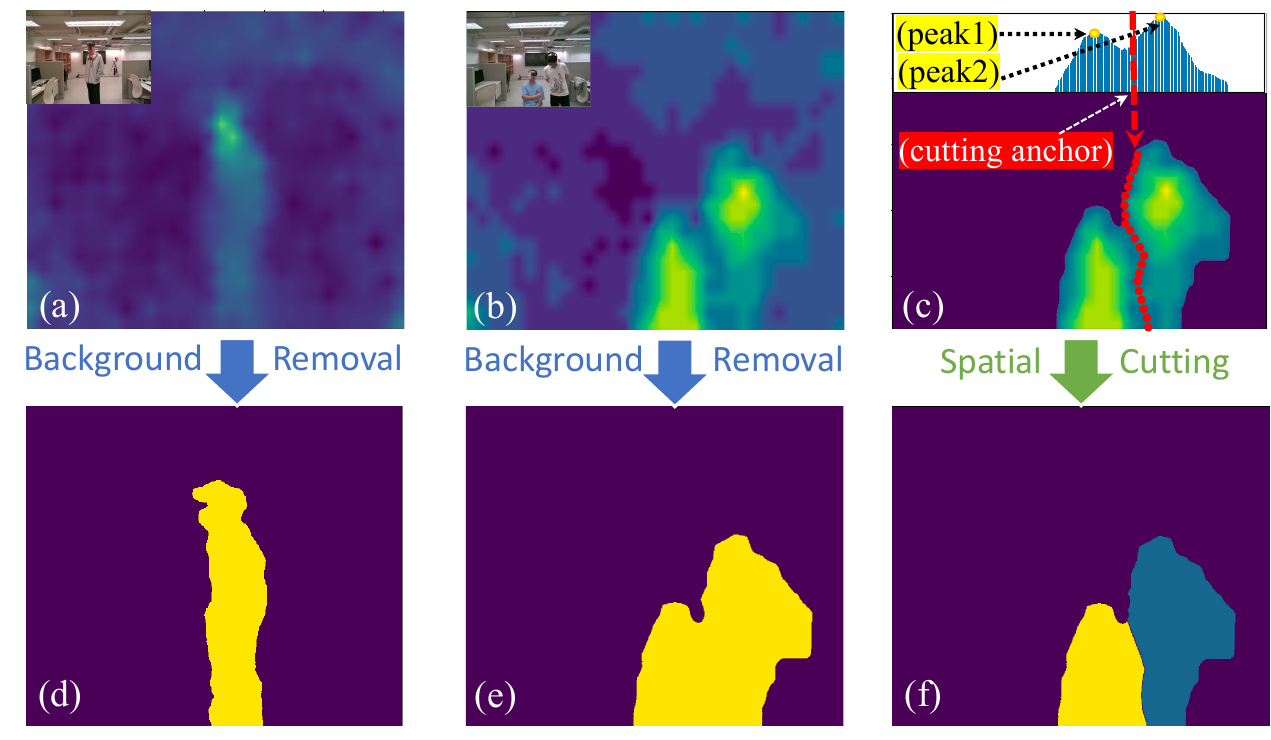}
    \vspace{-0.3in}
    \caption{Spatial detector. 
    }
    \label{fig:spatial-path}
  \end{minipage}
  \hfill
  \begin{minipage}{0.32\textwidth}
    \centering
    \includegraphics[width=1.0\textwidth]{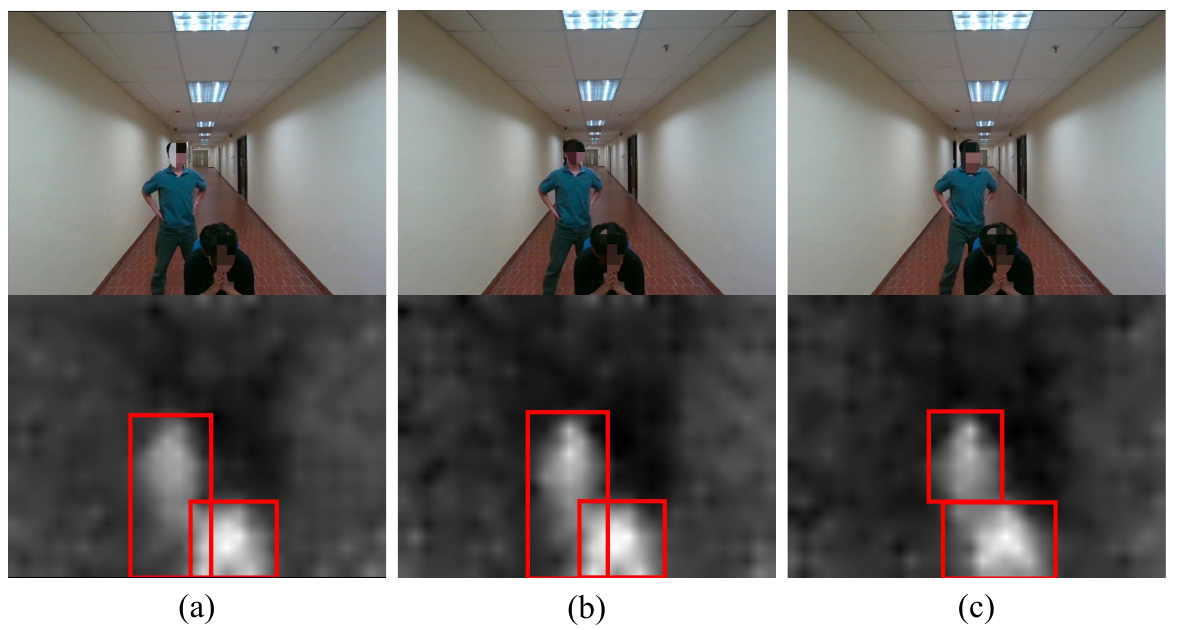}
    \vspace{-0.3in}
    \caption{Temporal Tracker.}
    \label{fig:temporal-path}
  \end{minipage}
\end{figure*}

\subsection{Preprocessing}
\label{ssec:preprocessing}
Before detection and range estimation, we preprocess raw thermal array maps in two steps. Firstly, we perform a bilinear interpolation with an expansion ratio $\theta$ to smooth the raw sensor readings and fill in missing pixels arising from the Melexis MLX90640 sensor's unique chessboard pattern, using a cut-off temperature of $37^\circ C$ to avoid interference from ambient high-temperature objects. The processed temperature map is directly used for the ranging model. Secondly, a copy of the map is normalized and rescaled to [0, 255] as a grayscale image for the spatiotemporal extractor.
\fig\ref{fig:preprocessing} illustrates the preprocessing steps, including (a) setup, (b) raw temperature map, (c) map with missing pixels filled in, (d) interpolated map with ratios of $\theta=10$ and (e) $\theta=20$, and (f) temperature map after removing out-of-range pixels.

\subsection{Thermal Array-based Human Detection}
\label{sec:roi-detection}
We need to detect multiple human targets from thermal array maps, a critical prerequisite for range estimation for each target. 
As mentioned in \S\ref{sec:intro}, several factors make the task challenging: 
1) The dynamic and non-uniform temperature pixels over a human body make it difficult to identify a complete mask for the body silhouette. 
2) The occupied pixels and their values for a distant user are too small to be recognized as a human target. 
3) Multiple users may create connected, overlapped, blurred pixels that are non-trivial to segment. 

The problem of multi-user separation appears to be similar to image segmentation in computer vision. However, the well-established CV approaches are not directly applicable here because the data are textureless temperature maps rather than visual RGB images and the resolution is extremely low as shown in our experiments \S\ref{sec:exp}. 
In \sysname, we propose a non-learning approach that does not require any training. 
The approach employs a \textit{spatial detector} and a \textit{temporal tracker} to jointly exploit the spatial distribution and temporal continuity of thermal frame pixels and then fuses the spatial and temporal detection for final results.

\head{Spatial Detector} 
The spatial detector extracts user regions from temperature distribution in a single thermal array map.
As shown in \fig\ref{fig:spatial-path}a, human target pixels often have higher temperatures indoors compared to other objects.
While a naive detection approach would apply a global threshold, this is impractical due to variations in thermal pixels across body parts. 
To reliably detect human-occupied regions and remove the background, we propose to use the adaptive Gaussian binary method \cite{opencv_library}, with the kernel size $s$ defined by the minimum size of a single user $\kappa$ and the expansion ratio $\theta$, \ie, $s = \kappa \times \theta$. 
As depicted in \fig\ref{fig:spatial-path}d and \fig\ref{fig:spatial-path}e, the above approach effectively removes the background and identifies human ROI, be it one user or multiple users. 
\rev{Note that all hyperparameter choices are detailed in \S\ref{sec:impl}.}

To facilitate multi-user separation, we analyze low temperature junctions and cut the multi-user ROI based on the following observation: 
Most commonly, as shown in \fig\ref{fig:spatial-path}, high-temperature pixels are concentrated around the human torso, irrespective of whether a user is standing, walking, or sitting upright.  
Consequently, a low-temperature depression is expected at the junction of two adjacent users.

To develop an effective separation method for horizontally connected users, we consider two constraints: 
1) Cutting paths must yield regions with similar temperature patterns, with the center being warmer, and 
2) Paths cannot pass through high-value areas representing the human body's core.
As illustrated in \fig\ref{fig:spatial-path}c, we generate a histogram representing column-wise sums across the foreground. Specifically, each column in the histogram, $C_i$, is the sum of all temperature values in that column: $C_i = \sum_{j=0}^{H}
{T_{(i,j)}}, i \in {0,1,2, \cdots, W}$, where $H$ and $W$ are the height and width of the preprocessed temperature map, respectively, and $T_{(i,j)}$ is the temperature value at the $(i,j)$ location.

The horizontal separation task is to find multiple thresholds, \ie, cutting anchors, $S_k \in \{1,2,3,\cdots,W\}$, that divide the histogram into multiple segments in a way that maximizes the variance between these segments. 
Drawing inspiration from Otsu's method \cite{ostu1979threshold}, the segmentation task is formulated as the following optimization problem and solved using the algorithm in \cite{liao2001fast}.
\begin{equation}
\begin{aligned}
\max_{S_1, S_2, \dots, S_{K-1}} \quad & \sigma_{B}^{2}(S_1, S_2, \dots, S_{K-1}) \\
\textrm{s.t.} \quad & 0 < S_1 < S_2 < \dots < S_{K-1}<W \\
            \quad & S_{0} = 0 \\
            \quad & S_{K} = W\\
\end{aligned}
\end{equation}
The inter-region variance $\sigma_{B}^{2}$ is defined as:
\begin{equation}
    \sigma_{B}^{2} = \sum_{k=1}^{K}{P_k(\mu_{k} - \mu_{G})^{2}},
    \label{eq-sigma}
\end{equation}
where $P_k = \frac{\sum_{i>S_{k-1}}^{i \leq S_{k}}{C_i}}{\sum_{i=0}^{W}{C_i}}$, $\mu_{k} = \frac{\sum_{i>S_{k-1}}^{i \leq S_{k}}{i \times C_i}}{\sum_{i>S_{k-1}}^{i \leq S_{k}}{C_i}}$, and $\mu_{G} = \sum_{k=1}^{K}{P_k \mu_k}$.

Although capable of multi-user separation, a straight vertical cut may pass through the user's body, breaching the second constraint.
To rectify this, we adjust each row's cutting point to a local temperature gradient minimum, creating an optimally curved path ( \fig\ref{fig:spatial-path}c).
By doing so, we segment all horizontally overlapped users as in \fig\ref{fig:spatial-path}f.

\head{Temporal Tracker} 
The spatial detector alone is insufficient, as it may produce false negatives for distant, low-temperature users or fail for vertically overlapping users, as shown in \fig\ref{fig:temporal-path}.
\rev{In addition, objects with temperatures similar to the human body may cause false alarms in the spatial detector.}
To facilitate the detection, we further incorporate a \textit{temporal tracker} to leverage the temporal correlation of continuous human movements. 
Since the thermal array readings are 2D elements, we can apply object tracking algorithms. 
However, most algorithms underperform due to the thermal array sensor's textureless and low-resolution features, including Camshift \cite{bradski1998computer}, dense optical flow \cite{farnebackTwoFrameMotionEstimation2003}, \etc. 
We find that the Kernelized Correlation Filter (KCF) tracking algorithm \cite{henriquesHighSpeedTrackingKernelized2015} is effective for thermal array data. 
As shown in \fig\ref{fig:temporal-path}, KCF employs a rectangular tracking box for each target, dynamically updated by searching for the most similar region in the vicinity of the next frame. 
To separate vertically overlapping users in \sysname, we calculate the horizontal and vertical overlap ratios for tracking boxes.
If the horizontal ratio exceeds a threshold and the vertical ratio remains below another threshold, we update the tracking boxes and perform vertical separation (as in \fig\ref{fig:temporal-path}c).
If both ratios exceed the thresholds, we will merge the two boxes as they are likely from the same user. 

\head{Spatial-Temporal Fusion} 
The KCF algorithm excels in ROI tracking but requires initial target regions, which is impractical. 
To deliver a practical ROI detector, we fuse the spatial detector and temporal tracker using the following steps: 
1) We remove random noise interference by filtering out non-overlapping regions in consecutive thermal array frames. 
2) For each region detected by the spatial detector, we form a contour bounding box. If it is unassociated with a tracking box, we initiate a new KCF tracker for it as it is likely a new user has entered. 
3) The bounding box with multiple overlapping tracking boxes is divided into vertically stacked boxes, assigning a single user per box. 
Finally, we obtain each user's ROI by extracting the corresponding bounding box region from the interpolated temperature map, as shown in \fig\ref{fig:infrage-overview}.

\begin{figure*}[t]
    \centering
    \includegraphics[width=0.9\textwidth]{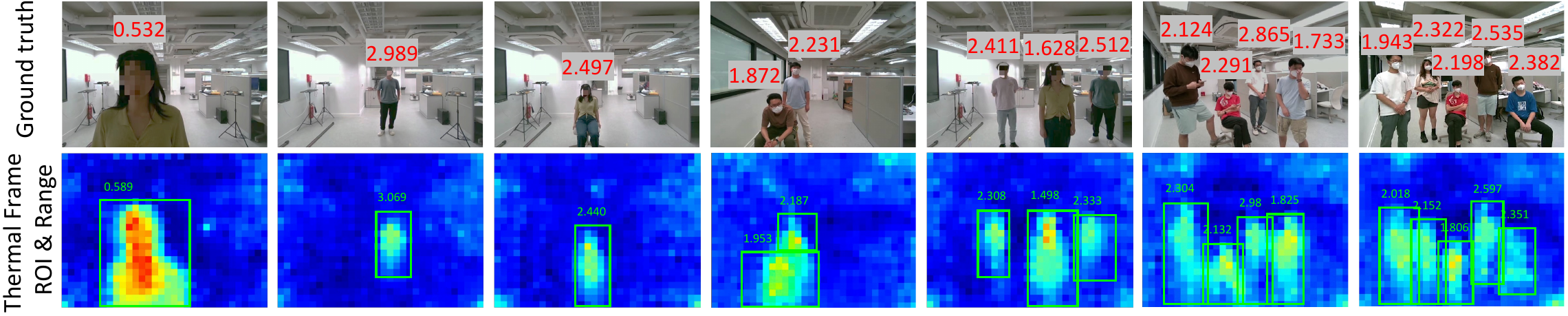}
    \vspace{-0.1in}
    \caption{\rev{Results of \sysname. {\rm The bottom row shows the thermal array maps with the detection and ranging results.}}}
    \label{fig:result-show}
\end{figure*}

\begin{figure}[b]
    \centering
    \includegraphics[width=0.44\textwidth]{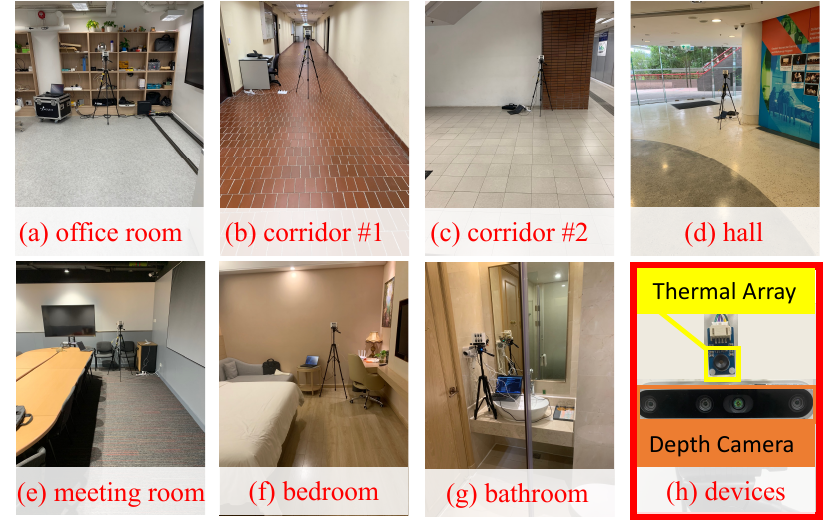}
    \vspace{-0.1in}
    \caption{Data collection environments and devices.}
    \label{fig:experiment-setup}
\end{figure}

\vspace{-0.05in}
\subsection{Range Estimation}
\rev{After obtaining the ROI for each user,  we take the individual ROI as input to the thermal ranging model designed in section \S\ref{sec:ranging}. 
By doing so, we obtain the range information for each human target as shown in \fig\ref{fig:result-show}.}

\begin{figure*}[t]
  \begin{minipage}{0.24\textwidth}
    \centering
    \includegraphics[width=1.0\textwidth]{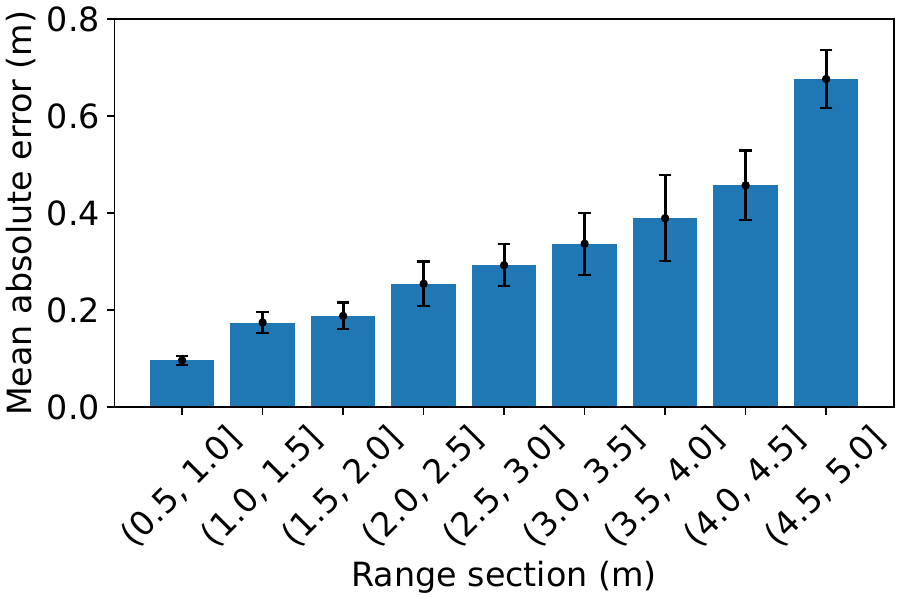}
    \vspace{-0.3in}
    \caption{\sysname ranging performance.}
    \label{fig:overall_range}
  \end{minipage}
  \hfill
  \begin{minipage}{0.24\textwidth}
    \centering
    \includegraphics[width=1.0\textwidth]{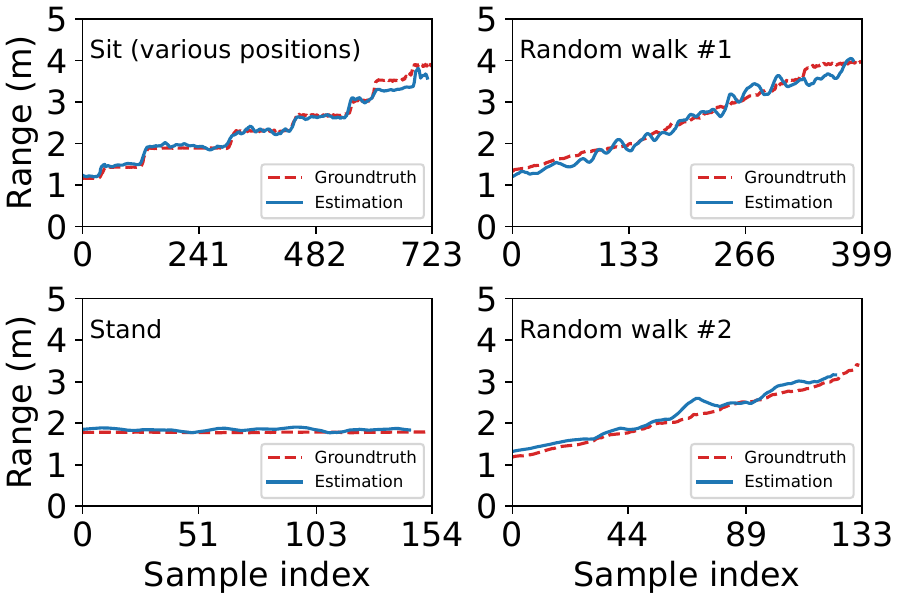}
    \vspace{-0.3in}
    \caption{Ranging estimation examples.}
    \label{fig:range_example}
  \end{minipage}
  \hfill
  \begin{minipage}{0.24\textwidth}
    \centering
    \includegraphics[width=1.0\textwidth]{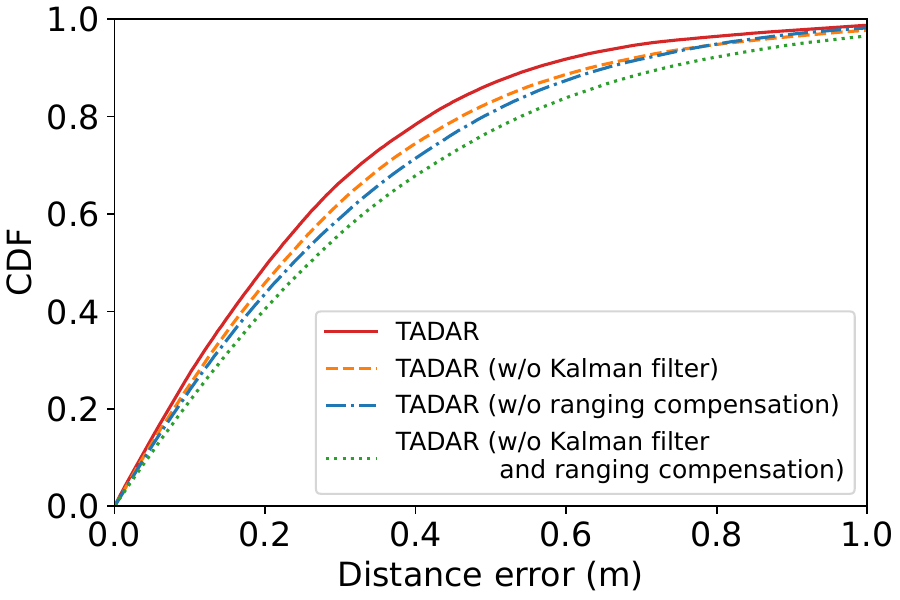}
    \vspace{-0.3in}
    \caption{Ranging estimation CDF.}
    \label{fig:overall_range_cdf}
  \end{minipage}
  \hfill
  \begin{minipage}{0.24\textwidth}
    \centering
    \includegraphics[width=1.0\textwidth]{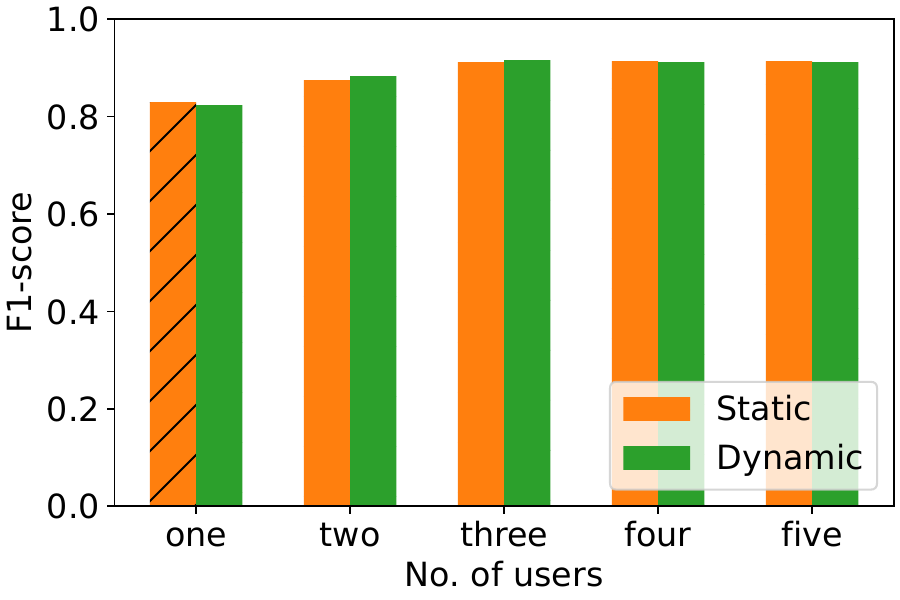}
    \vspace{-0.3in}
    \caption{\rev{\sysname detection performance}}
    \label{fig:overall_detection}
  \end{minipage}
\end{figure*}

\vspace{-0.05in}
\section{Implementation}
\label{sec:impl}

\head{Parameters}
Our default setting uses a pooled ROI size of $2 \times 4$, resulting in a feature dimension of $n=8$.
In preprocessing (\S\ref{ssec:preprocessing}), we apply bilinear interpolation with an expansion ratio $\theta=400$, expanding each dimension by 20 times. 
Considering the resolution and FOV of our thermal array sensor, each element of the temperature map corresponds to approximately $44.6 \times 31.9 cm^2$ at a 5-meter distance. Therefore, we set the minimum size for a single user's region, $\kappa$, in the temperature map to 5, based on the typical body size.

\head{Devices}
Our prototype uses the MLX90640BAA thermal array sensor, which is a low-cost, lightweight device with $32 \times 24$ elements and a $110\degree \times 75\degree$ FOV. The sensor is configured with a 16 Hz refresh rate in our system, and the emissivity is set to 1. The entire system operates on a Raspberry Pi CM4.

\begin{figure}[t]
    \subfloat[Single-user ranging.]{%
    \label{subfig:baseline_range}
      \includegraphics[width=0.23\textwidth]{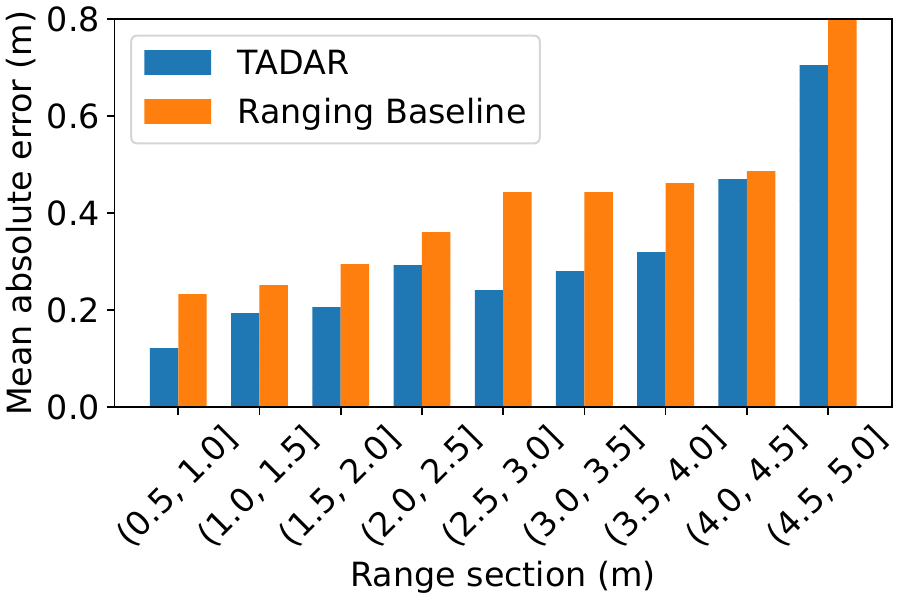}
	  }
    \subfloat[Uers detection.]{%
    \label{subfig:baseline_detection}
      \includegraphics[width=0.23\textwidth]{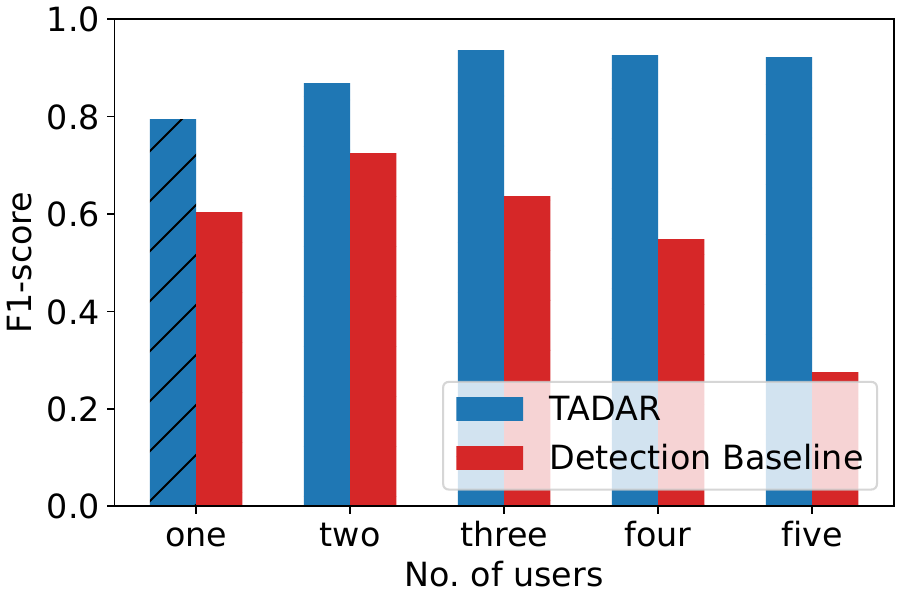}
	  }
   \vspace{-0.6\baselineskip}
    \caption{\rev{\sysname performance compared to baselines.}}
	\label{fig:baselines}
\end{figure}


\section{Experiments}
\label{sec:exp}
\subsection{Dataset, Baselines, and Metrics}
\head{Dataset} We collected a dataset with over 70,000 samples across seven indoor environments, as shown in \fig\ref{fig:experiment-setup}.
Our primary focus is on indoor settings due to reduced privacy concerns in outdoor public spaces. 
Seven participants (six males and one female) were recruited with IRB approval from our institution.
A RealSense D455 depth camera captured depth and RGB video, providing ground truth.
To train the HGB algorithm, we use part of the data in the office from two users for training, and keep all the rest for testing.

\head{Baselines} \sysname is the first thermal array-based multi-user detection and ranging system. To demonstrate its advantages, 
two baselines are considered in our experiments. 1) Ranging Baseline \cite{naserHumanDistanceEstimation2021} focuses on the thermal array-based single-user ranging, which is motivated by the “smaller farther, bigger closer” effect, and mainly relies on the number/size of user-occupied pixels in the bottom lines for ranging and only touched on temperatures as supplementary features. Consequently, \cite{naserHumanDistanceEstimation2021} is limited to single-user strictly in front of the device, degrading with various body sizes and failing in multi-user or blocked bottom lines. 2) Detection Baseline \cite{naserAdaptiveThermalSensor2021} performs occupancy estimation with thermal-array sensor, where the U-Net \cite{ronnebergerUNetConvolutionalNetworks2015a} is adopted for human segmentation.

\head{Metrics} 
\sysname was assessed using mean absolute error and mean relative error, \ie, mean absolute error divided by the actual range, with successful ROI detection determined by an intersection over union (IoU) greater than 0.5 between the detected ROI and ground truth bounding box.

\begin{figure*}[t]
  \begin{minipage}{0.24\textwidth}
    \centering
      \includegraphics[width=1.0\textwidth]{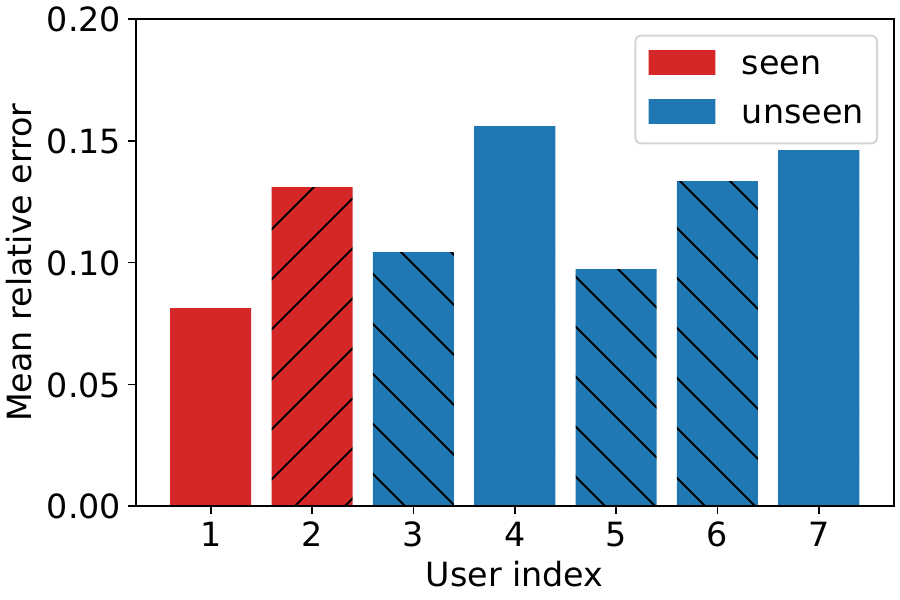}    
    \vspace{-0.3in}
    \caption{Performance on different users.}
	\label{fig:cross-user-range}
  \end{minipage}
  \hfill
  \begin{minipage}{0.24\textwidth}
    \centering
    \includegraphics[width=1.0\textwidth]{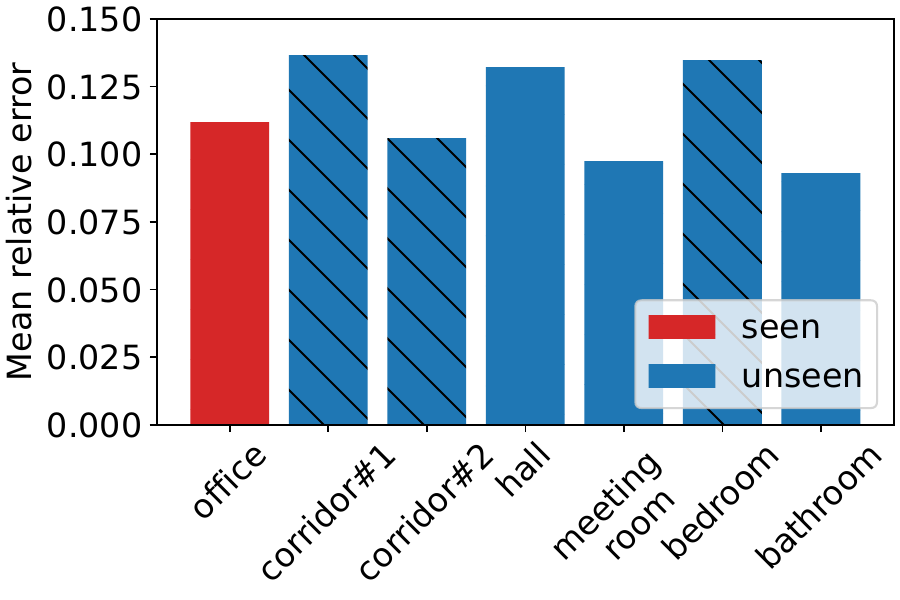}
    \vspace{-0.3in}
    \caption{Performance across environments.}
    \label{fig:cross-env-range}
  \end{minipage}
  \hfill
  \begin{minipage}{0.24\textwidth}
    \centering
    \includegraphics[width=1.0\textwidth]{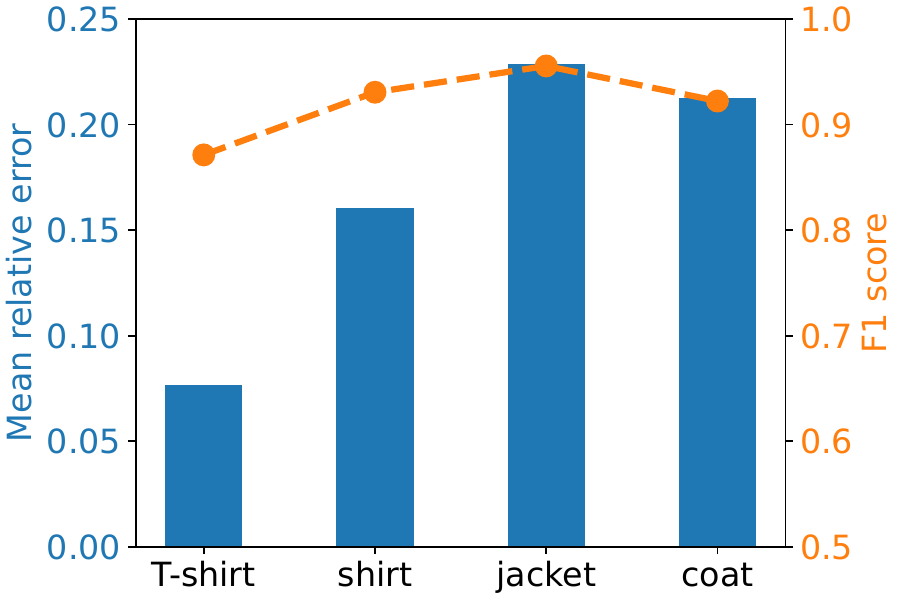}
    \vspace{-0.3in}
    \caption{Performance with various clothes}
    \label{fig:cross-clothes-range}
  \end{minipage}
  \hfill
  \begin{minipage}{0.24\textwidth}
    \centering
    \includegraphics[width=1.0\textwidth]{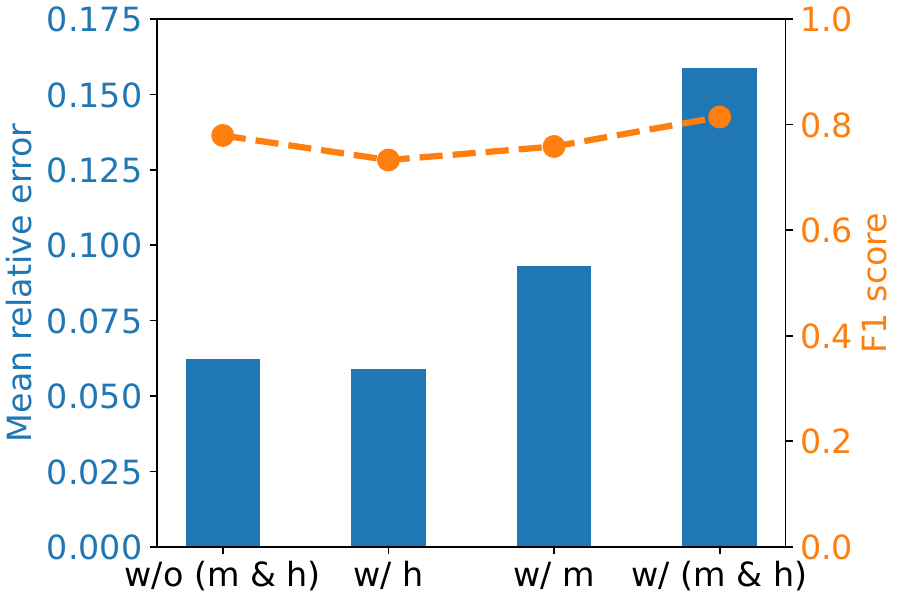}
    \vspace{-0.3in}
    \caption{Impact of the mask and hat.}
    \label{fig:cross-maskhat-range}
  \end{minipage}
\end{figure*}

\begin{figure*}[t]
  \begin{minipage}{0.3\textwidth}
    \centering
    \subfloat[Orientations.]{%
    \label{subfig:cross-orientation-case}
      \includegraphics[width=0.35\textwidth]{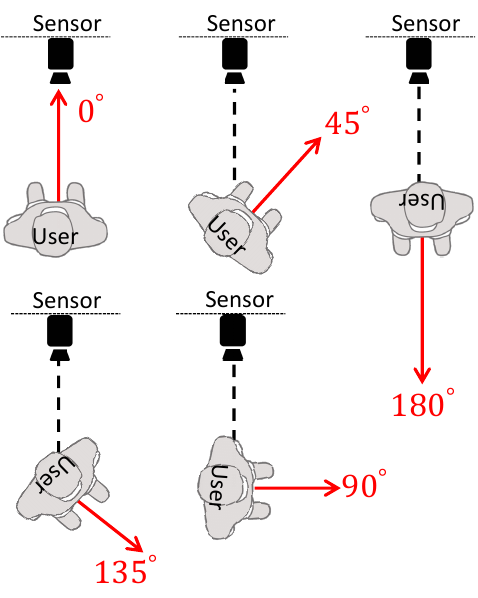}
	  }
    \subfloat[Performance.]{%
    \label{subfig:cross-orientation-range}
      \includegraphics[width=0.65\textwidth]{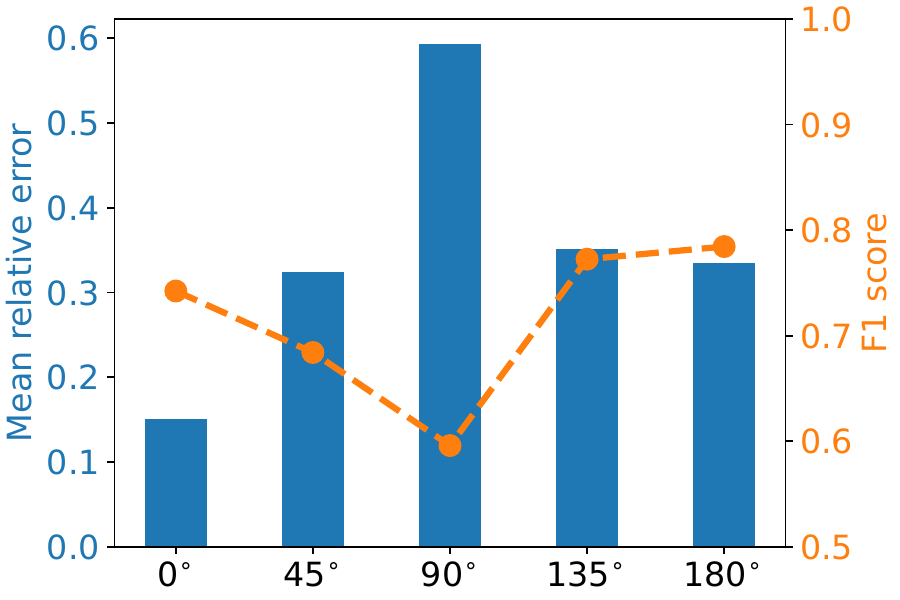}
	  }
   \vspace{-0.6\baselineskip}
    \caption{The performance w.r.t. user orientations.}
	\label{fig:cross-orientation}
  \end{minipage}
  \hspace{0.02\textwidth}
  \begin{minipage}{0.26\textwidth}
    \centering
    \subfloat[Angles.]{%
    \label{subfig:cross-incidentangle-case}
        \includegraphics[width=0.40\textwidth]{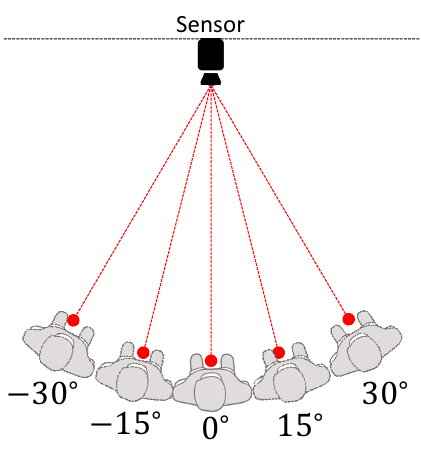}
	  }
    \subfloat[Performance.]{%
    \label{subfig:cross-incidentangle-range}
    \includegraphics[width=0.60\textwidth]{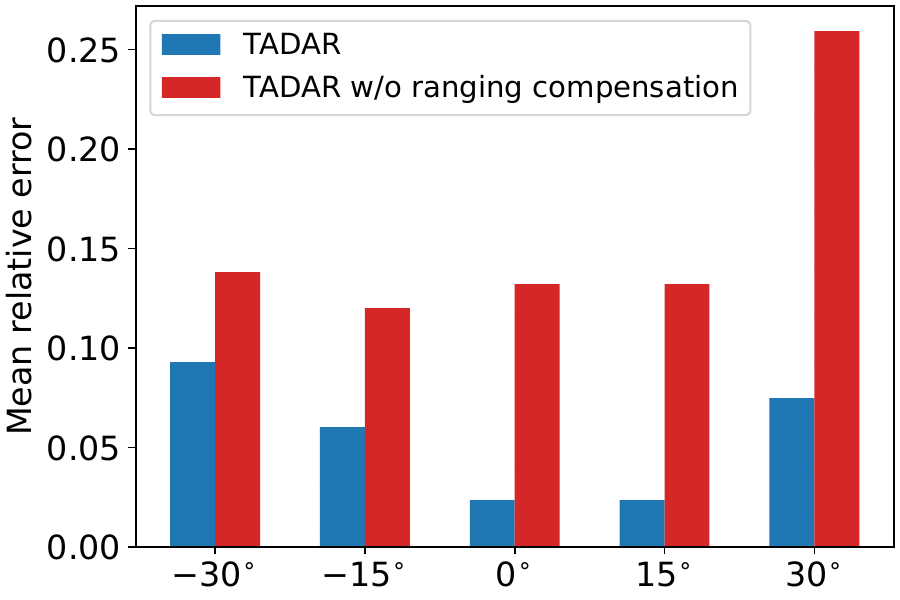}
   }
   \vspace{-0.4\baselineskip}
    \caption{The performance w.r.t. incident angles.}
	\label{fig:cross-incidentangle}
  \end{minipage}
  \hspace{0.02\textwidth}
  \begin{minipage}{0.33\textwidth}
    \centering
    \subfloat[Ambient objects.]{%
    \label{subfig:impact-ambientobjects-case}
    \raisebox{10pt}{
      \includegraphics[width=0.4\textwidth]{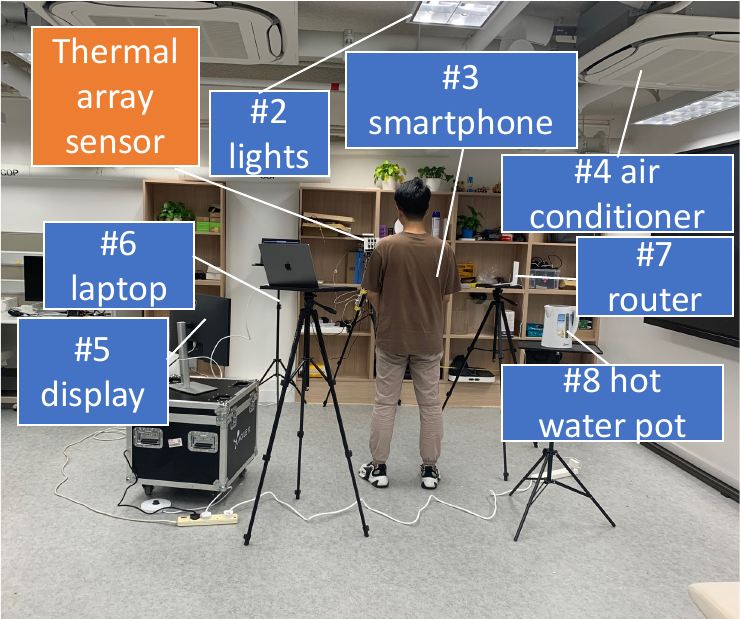}
	  }
   }
    \subfloat[Performances.]{%
    \label{subfig:cross-ambientobjects-range}
      \includegraphics[width=0.6\textwidth]{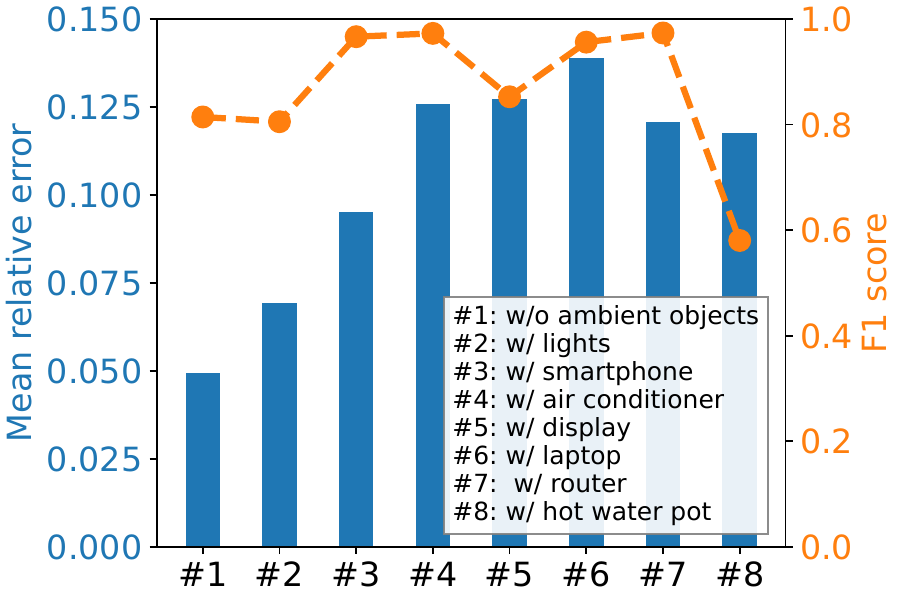}
	  }
   \vspace{-0.6\baselineskip}
    \caption{Impact of the ambient objects.}
	\label{fig:impact-ambient-objects}
  \end{minipage}
\end{figure*}

\vspace{-0.15in}
\subsection{Overall Performance}
\head{Human Ranging} 
The ranging performance, depicted in  \fig\ref{fig:overall_range}, correlates with the subject's absolute distance.
Optimal results occur within 3 meters, with a mean error of 20.1 cm, increasing to 39.4 cm between 3 and 4.5 meters. 
Considering the body size, the accuracy is sufficient to support many downstream sensing applications, such as intrusion detection and indoor human tracking. 
Specifically, we notice that the ranging error increases with respect to the user's actual range.
Since the thermal array sensor has extremely low resolution, a distant user only occupies a few pixels in the temperature map. This reduces signal-to-noise ratio (SNR) and increases variances, leading to larger ranging errors.
Moreover, \fig\ref{fig:range_example} shows a time series of range estimates for activities like standing and walking, and  \fig\ref{fig:overall_range_cdf} presents the cumulative distribution function (CDF) of \sysname's ranging error, with the 80th percentile at 41.7 cm. 
The figure also indicates that accounting for horizontal bias and temporal information improves performance.

\head{Human Detection} 
The human detection performance of \sysname is assessed in both dynamic (e.g., walking) and static (e.g., standing or sitting) scenarios, as shown in \fig\ref{fig:overall_detection}. 
Surprisingly, the F1-score for single-user detection is lower than in multi-user situations.
This is because \sysname's spatial detector relies on the temperature distribution difference between the background and the target region. 
Sensor noise will amplify this difference when focusing on the cooler background, leading to more false alarms and a lower F1 score in single-user cases. 
Comparing dynamic and static human detection, the detector shows similar performance, indicating its robustness in handling free-walking motions.

\head{Baseline Comparison} 
Our system surpasses both ranging and detection baselines. 
\rev{It outperforms the range estimator of the ranging baseline \cite{naserHumanDistanceEstimation2021} by 25.01\% in MAE in its favored single-user scenarios (\fig\ref{subfig:baseline_range}) and exceeds the detection baseline \cite{naserAdaptiveThermalSensor2021} in all scenarios, with an average gain of 36.76\% in F1-score (\fig\ref{subfig:baseline_detection}).}

\head{System latency}
Analyzed on a Raspberry Pi CM4, the average processing time of \sysname is 1.338s per frame, suitable for real-time sensing at a low update rate. Detection takes 1.328s, while ranging requires only 0.016s. A C++ version is planned for faster processing.

\vspace{-0.15in}
\subsection{Benchmark Evaluation}
\label{ssec:exp_be}
\head{Cross-user} 
We evaluate \sysname across seven users, training with data from two and testing with the others. For each user, we collect data at varying distances (0.5 m to 5 m) from the sensor. \fig\ref{fig:cross-user-range} shows minimal differences among users, indicating strong generalizability across users.

\head{Cross-environment} 
The HGB model in \sysname, initially trained on the data from an office, is tested in various environments including corridors, halls, bathrooms, \etc. As shown in \fig\ref{fig:cross-env-range}, \sysname maintained consistent performance, with mean relative errors ranging from 9.3\% to 13.7\%.

\head{Impact of clothing}
Clothing can affect user thermal radiation and thus sensing performance. \fig\ref{fig:cross-clothes-range} demonstrates that \sysname performs well with users wearing T-shirts or shirts. Even with jackets or coats, effectiveness is maintained if the face and neck areas remain exposed. Overall, \sysname supports sensing applications regardless of the clothing type.

\head{Impact of mask and hat} 
\sysname relies on facial and neck temperatures. \fig\ref{fig:cross-maskhat-range} shows the results with different mask and hat combinations. Detection performance is consistent, but ranging accuracy decreases when a mask or hat is worn, with masks having a greater impact due to significant facial coverage. However, performance is still acceptable for many real-world applications, as wearing both a mask and a hat is uncommon in homes and offices.

\head{Cross-orientation evaluation} 
\sysname is most effective when detecting bare skin temperature, like facial temperature. We evaluated its performance at different user orientations relative to the thermal array sensor, across five conditions (\fig\ref{subfig:cross-orientation-case}). Results (\fig\ref{subfig:cross-orientation-range}) show that \sysname performs best when the user's face is fully exposed to the sensor. Performance decreases at a $90^{\circ}$ orientation due to a reduced visible body area, which leads to a loss of range information and a lower detection rate. However, the real-life impact is minimal as users rarely maintain at $90^{\circ}$ orientation and multi-sensor fusion can enhance performance.

\head{Impact of incident angle} 
We evaluated the influence of horizontal bias and the effectiveness of our proposed ranging compensation method. We analyzed performance at various incident angles with users positioned between 1.5 m and 2 m (\fig\ref{subfig:cross-incidentangle-case}) and compared ranging accuracy with and without compensation. As depicted in \fig\ref{subfig:cross-incidentangle-range}, the ranging errors decrease across all incident angles with ranging compensation. Additionally, errors increase with larger incident angles due to random noise in thermal array map pixels, which becomes more pronounced near the border.

\head{Impact of ambient objects}
Indoor objects emit thermal radiation, potentially interfering with user radiation. To assess this, we conduct experiments with common objects as shown in \fig\ref{subfig:impact-ambientobjects-case}. 
Evaluating each object's influence, we remove or turn off other objects, placing the target object at various positions within the thermal array's FOV.
As shown in \fig\ref{subfig:cross-ambientobjects-range}, the light tube has minimal impact on ranging performance, while other objects have similar effects. 
\sysname maintains good ranging performance despite interference, but detection F1 score decreases with the hot water bottle present. 
Nevertheless, the recall rate remains high at 95.2\%, with good ranging accuracy, still satisfactory for most applications.
Ranging for non-human objects and recognition of non-human objects are considered as our future work.

\head{Impact of light condition} 
We evaluate \sysname under different lighting conditions (bright, normal, dim, and dark), which shows consistent performance with ranging errors (5.6\% to 7.3\%) and detection F1 scores (79.1\% to 82.1\%).

\vspace{-0.15in}
\subsection{Case study}

\begin{figure}[t]
  \begin{minipage}{0.23\textwidth}
    \centering
    \subfloat[Mapping in-frame height $h$ to real height $H$ with range $r$.]{%
    \label{subfig:fall-detection-principle}
      \includegraphics[width=0.9\textwidth]{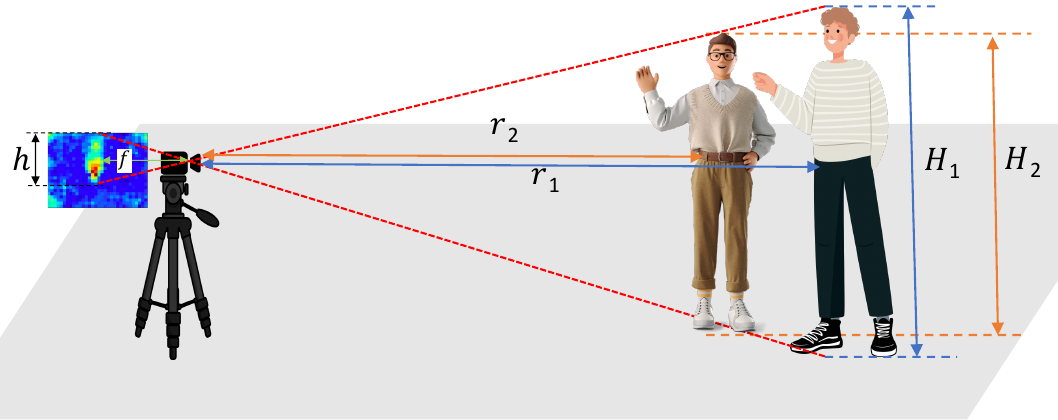}
	  }
 	  \vfill
    \vspace{-0.8\baselineskip}
    \subfloat[Real/in-frame height gradients w.r.t. user activities.]{%
    \label{subfig:fall-detection-curve}
      \includegraphics[width=0.9\textwidth]{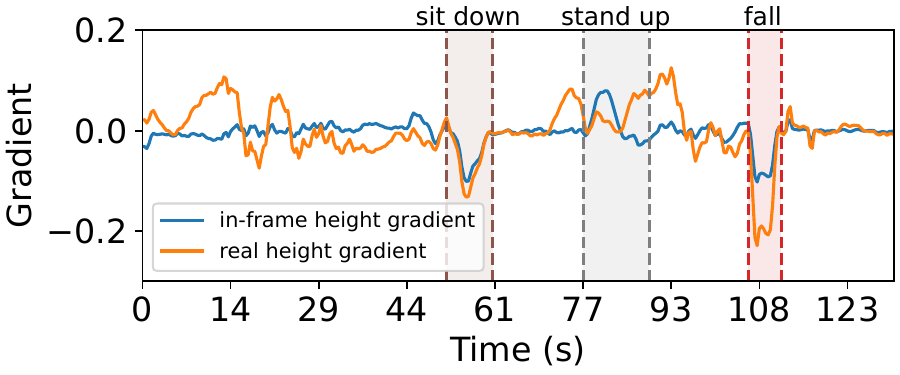}
	  }
    \caption{\sysname for fall detection.}
	\label{fig:fall-detection}
  \end{minipage}
  \hfill
  \begin{minipage}{0.23\textwidth}
    \centering
    \vspace{0.8\baselineskip}
     \raisebox{7pt}{
    \includegraphics[width=0.95 \textwidth]{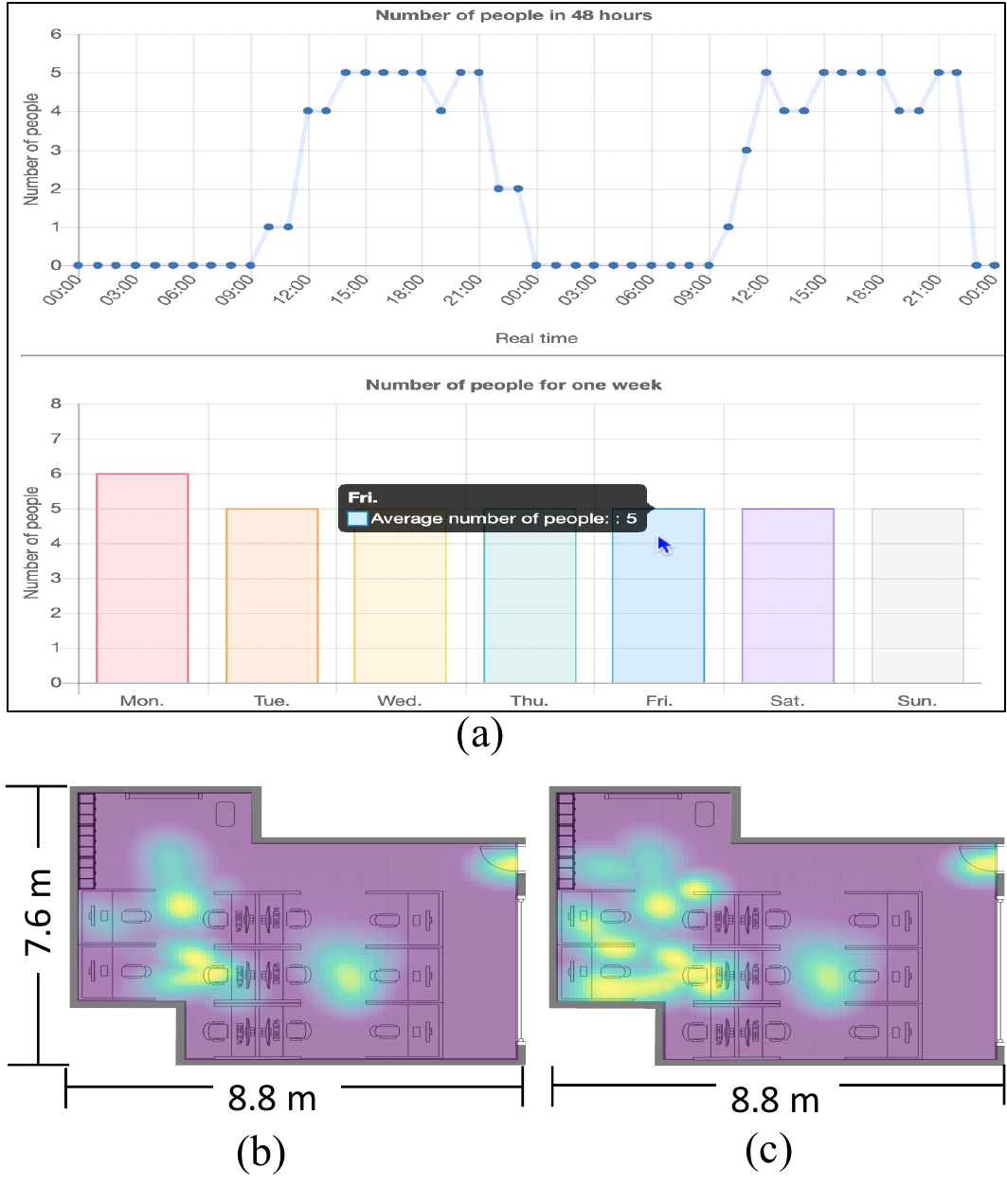}
    }
    \vspace{-0.4 in}
    \caption{\sysname for occupancy monitoring. 
    }
    \label{fig:human-monitoring}
  \end{minipage}
\end{figure}

\head{Fall detection} Owing to its spatial resolution and privacy protection, \sysname holds promise for fall detection. 
The key indicator of a fall is a rapid change in the user's height.
Empowering thermal arrays with range enables the calculation of a user's actual height and thus allows us to build a fall detection system.
As shown in \fig\ref{subfig:fall-detection-principle}, the in-frame height $h$ does not directly correspond to the user's actual height.
\fig\ref{subfig:fall-detection-curve} shows that the gradient of the in-frame height cannot distinguish between sitting down and falling due to overlapping gradient values.
However, with the estimated range $r$, we can calculate the real height $H$ as $H \approx \frac{h \times r}{f}$, where $f$ is the thermal array's focal length.
By utilizing the real height and thresholding the gradient values, we can accurately detect falls and reduce false alarms from other activities. 
In an experiment involving 22 falls with five users, we achieved a 95.5\% detection rate without false alarms, compared to an 86.4\% detection rate with one false alarm when using only the in-frame height without range. Although this initial design requires refinement for practical use, it demonstrates the feasibility of fall detection with \sysname as a key enabler.

\head{Occupancy monitoring} 
\rev{With the range information, the thermal array sensor can be easily deployed on the wall, rather than on the ceiling, to perform long-term occupancy monitoring (\ie, user counting). 
We build a prototype system based on \sysname and mounted it on a wall at approximately 1.6 m height in the office for over one week.} 
As depicted in \fig\ref{fig:human-monitoring} (a), the demo system tracks the number of people within the office. 
Furthermore, by combining ranging results with the detected in-frame position, the system can generate occupancy heatmaps to indicate the active areas, as depicted in \fig\ref{fig:human-monitoring} (b-c). 
This study highlights that \sysname paves the way for easy-to-install occupancy system for many applications, \eg, intrusion detection and building automation.

\vspace{-0.15in}
\section{Related Works}
\label{sec:related_works}

\head{Human Sensing}
Human sensing has been a vast research area that attracts numerous and increasing research efforts, with many innovative applications, such as fall detection \cite{huDeFallEnvironmentIndependentPassive2021}, gesture control \cite{wang2016device}, fitness tracking \cite{wuGaitWayMonitoringRecognizing2021}, etc. 
Various technologies have been explored to enable contact-free human sensing, including cameras \cite{ lagaSurveyDeepLearning2022}, thermal cameras \cite{wangRGBInfraredCrossModalityPerson2019}, WiFi \cite{pallaprolu2022wiffract,qianWidar2PassiveHuman2018,kong2021multiauth}, sound \cite{zhang2023vecare, maoDeepRangeAcousticRanging2020}, mmWave radars \cite{kongM3TrackMmwavebasedMultiuser2022, zhang2020mmeye}, PIR sensors \cite{guanDailyActivityRecognition2019}, \etc. 
Many of these works also focus on human ranging \cite{maoDeepRangeAcousticRanging2020,lagaSurveyDeepLearning2022}. 
However, these technologies generally confront a conflict between imaging resolution and privacy protection. 
In contrast, thermal array sensors offer a good balance and promise an attractive sensing modality. 

\head{Thermal Array Sensing}
Thermal array sensors have attracted increasing attention, showing great potential in many sensing applications, such as fall detection \cite{tatenoPrivacyPreservedFallDetection2020}, occupancy estimation \cite{chiduralaDetectionMovingObjects2022, chiduralaOccupancyEstimationUsing2021,naserAdaptiveThermalSensor2021}, and human monitoring \cite{naserMultipleThermalSensor2022}. 
In \cite{naserAdaptiveThermalSensor2021}, the authors employ the U-Net \cite{ronnebergerUNetConvolutionalNetworks2015a} to perform human segmentation using a thermal array installed on the wall. 
To gain the range information, 
\cite{naserHumanDistanceEstimation2021} relies heavily on user-occupied pixels in the bottom lines for ranging, effective for only one user strictly in front and failed with varying body sizes or in multi-user settings. 
In summary, lacking range information, existing thermal array sensing approaches are mostly limited to using 2D temperature maps as regular images, and sensors often need to be installed on the ceiling for sensing \cite{naserMultipleThermalSensor2022,chiduralaDetectionMovingObjects2022,tatenoPrivacyPreservedFallDetection2020, chiduralaOccupancyEstimationUsing2021}. 
\sysname enables the missing range information and paves the way for ubiquitous thermal array sensing. 

\vspace{-0.15in}
\section{Conclusions}
\label{sec:conclusion}
We introduce thermal array sensors as an emerging modality for ubiquitous human sensing that effectively balances imaging resolution and privacy. 
We present \sysname, the first multi-user thermal array-based detection and ranging system that converts thermal pixels into ranges. 
We build a real-time prototype using a commodity thermal array sensor and conduct extensive experiments in real-world environments, demonstrating the remarkable performance of \sysname. 
\rev{By making the inherently missing range information available for thermal array sensing, we believe \sysname underpins ubiquitous thermal array sensing and opens up a new sensing direction beyond wireless, acoustic, and light-based methods.}

\section{Acknowledgments}
We thank our shepherd Joerg Widmer and the anonymous reviewers for their valuable feedback on this paper. 
This paper is partly supported by the NSFC under grant No. 62222216, the Hong Kong RGC ECS under grant 27204522, and RIF under grant R5060-19.